\documentclass[11pt]{article}
\usepackage[a4paper,margin=25mm]{geometry}
\usepackage{amsmath,amssymb,bm,booktabs,graphicx,xcolor}
\usepackage[round,authoryear]{natbib}
\newcommand{\utau}{u_\tau}

\usepackage{array}
\providecommand{\Rey}{\mbox{\textit{Re}}}
\newcommand{\Ret}{\Rey_\tau}
\newcommand{\lp}{\ell^+}
\newcommand{\lameff}{\lambda_{\mathrm{eff}}}

\newcommand{\CT}{C_T}
\newcommand{\CU}{C_U}
\usepackage{url}
\usepackage[hidelinks]{hyperref}
\usepackage{cleveref}
\usepackage{doi}  

\title{Tracking the multiscale celerity of turbulent wall footprints}
\author{M. Miozzi\\[4pt]
\small INM-CNR, Institute of Marine Engineering, National Research Council,\\
\small Via di Vallerano 139, 00128 Rome, Italy\\
\small\texttt{massimo.miozzi@cnr.it}\\
\small ORCID: \href{https://orcid.org/0000-0002-0031-9594}{0000-0002-0031-9594}}
\date{}

\begin{document}
\maketitle

\begin{abstract}
The celerity of a wall footprint, the speed at which a temperature, stress or pressure feature moves on the wall, can be measured scale by scale as a pattern displacement. We present a class of Lagrangian trackers that measure it as a field. Each image is split into several bands at fixed wavelength, and every feature is followed from one image to the next on every band. The result is a celerity per feature, per scale and per instant, on the same wall points for all scales. 
We adopt two formulations of the tracker: one holds constant
the band intensity, the other the local phase of a quadrature bank. Their outcomes agree within one per cent on a DNS of a turbulent channel. On benchmarks, trackers resolve a celerity that changes with scale or along the wall, and their performances have been checked against a reference correlation-peak code. The benchmarks also show that additive noise can bias the measurement only through what the two frames share, so the interval between them, rather than the image quality, controls that bias. The trackers are applied to the stress and the pressure of the turbulent channel, and to the temperature pattern recorded by TSP in a turbulent boundary layer. The celerity is thus obtained as a local quantity, which can be mapped, followed along a path and compared with the motions above the wall: every point, carried by the broadband motion of the pattern, is a Lagrangian probe that carries all the scales.
\end{abstract}

\noindent\textbf{Keywords:} Wall footprints, Celerity, Feature tracking, Taylor hypothesis, Temperature-sensitive paint
 
\section{Introduction}\label{sec:intro}
Optical diagnostics are widely used in experimental fluid mechanics. Particle Image Velocimetry (PIV) is the standard technique for spatially resolved velocity fields \citep{raffel2018}, and has grown from planar to time-resolved and volumetric measurements \citep{rohacs2023,wang2025}. Particle Tracking Velocimetry (PTV) follows individual tracers rather than window displacements, and avoids part of the spatial averaging of correlation \citep{dabiri2019}. Its modern form, Lagrangian Particle Tracking (LPT), reconstructs dense trajectories in three-dimensional, time-resolved volumes \citep{schanz2016,schroeder2023}, at Reynolds numbers previously out of reach \citep{kuchler2024}.
The progression from PIV to PTV and LPT represents a transition from measuring velocity fields to identifying the motions that carry transport. 

In recent years, a renewed interest has developed in near-wall turbulence, in particular for what concerns the propagation celerity of wall footprints at a given scale. Since \citet{eckelmann1974} and \citet{kim1993}, the celerity of each scale has been a physically meaningful quantity, because it can be linked to the turbulent motions that carry that scale. Close to the wall, systematic measurements are still a challenging task. Strong velocity gradients in narrow regions are hard to resolve with seeded tracers. Where an adverse pressure gradient separates the boundary layer, the reverse flow makes a uniform distribution of the tracers even harder to achieve.

Temperature- and Pressure-Sensitive Paint (TSP and PSP) capture maps of wall temperature and pressure at high spatial and temporal resolution. They look at the turbulence from the wall itself, and need no material seeding, because the pattern to track is the texture of the wall footprint. Transitional and turbulent flows provide such a pattern, while a laminar flow offers none. This specifies the range of application of the proposed technique. Transitional and turbulent flow conditions allow TSP and PSP to make it possible to track these footprints scale by scale. Here we investigate how the celerity of each scale can be obtained by tracking the temperature, stress and pressure wall patterns, and how far the measurement can be trusted.

The celerity of a surface pattern from TSP was first used as a skin-friction diagnostic in  \citet{miozzi2019}. The initial correlation-based approach moved to a more refined one, where the celerity was then taken from the space and time derivatives of the  sequence of TSP maps as the velocity that best satisfies a convection equation averaged in time, following a weak form of Taylor's hypothesis \citep{delalamo2009,miozzi2020}, WTH in the following. Tied to the friction velocity, it gave the wall stress with no thermal calibration, by using the relationships proposed by \cite{hetsroni2004} to move from the celerity of the fluctuations of temperature to the velocity ones. The method was applied, at a single broadband scale, to transonic flow \citep{costantini2021}, and compared with Shake-The-Box (STB) velocity data  around a wall-mounted cube \citep{miozzi2024} and across a backward facing ramp \citep{costantini2026ramp}. Its link with the convection of enstrophy was investigated by \citet{liu2024}. The WTH returns, pixel by pixel, a two-component Eulerian celerity field averaged in time. 

The trackers adopted in this paper descend from YATS, a correlation--based feature tracker built on the Lucas--Kanade solver and introduced in PIV by \citet{miozzi2004particle} \citep[see also][]{miozzi2005,miozzi2008,stanislas2008}. YATS has been applied to a range of flows, among them the drag reduction by microbubbles in a turbulent boundary layer \citep{jacob2010}, a free-surface jet with Coanda effect \citep{miozzi2010experimental}, the propagation of pressure waves and the distribution of loads in wave slamming \citep{lugni2010}, and a wavy boundary layer over a permeable wall \citep{Miozzi2015}. Unlike the classical PIV approaches, which look for the maximum of the cross-correlation \citep{adrian2007}, this tracker takes as its best measure of correlation the minimum of the Sum of Squared Differences (SSD) between the intensities of the interrogation windows, iteratively warped by a prescribed linear deformation and resampled at sub-pixel displacements. Being the square of a binomial difference, the SSD is composed of the cross-correlation with the energies of the two windows added; the energy of the warped window changes with the displacement. 

A tracker of this kind rests on the Brightness Constancy Constraint (BCC): a pattern keeps its brightness from one frame to the next. A TSP footprint does not, since the temperature fluctuation that marks it changes as it moves. This is the main reason why our work on TSP followed the WTH, which imposes the same equation in the time mean over the whole sequence rather than on each pair. 

Before a TSP footprint can be tracked pair by pair, one must know how far the change in intensity biases the tracker.
To this end we developed YATTS (Yet Another Temperature Tracking Software). It keeps the intensity solver of YATS, adds a solver based on Phase Constancy Constraint \citep{fleet1990}, independent of a uniform change in the image intensity, and measures several bands on the same wall points. The tests show that, on a benchmark built from frames of the turbulent channel (\S\ref{sec:fields}), a uniform intensity decay biases neither solver, while a decay linked to the local stress biases mainly the coarsest band of both, through the gradient of its envelope (\S\ref{sec:mcdecay}). Tracking and the WTH are applied to the same datasets and compared in \S\ref{sec:wt-compare}. 

The celerity of wall footprint resolved by scale has already been measured, as a mean per wavenumber: in a channel DNS from local time derivatives \citep{delalamo2009}, in a boundary layer at $Re_\theta = 13\,000$ \citep{renard2015}, from the linearised equations about the mean profile \citep{Liu_Gayme_2020}, and, with its variability, from measured space--time spectra \citep{cui2024variability}. \citet{foroozan2025instantaneous} measured the convection of the infrared footprint of the wall heat transfer from space--time correlations: the larger streaks travel at about the free-stream speed, the smaller features at about $10\,\utau$, close to the fine bands of the stress in the channel (\S\ref{sec:dns}). Other methods give part of this information. \citet{lozano2014time} followed coherent structures in time, in the volume of a channel DNS; the structures are not separated by scale and are not wall footprints. Optical flow from the local phase of band-pass filters \citep{fleet1990,gautama2002}, or from wavelets \citep{derian2013}, returns one velocity per band, but as an Eulerian field, without trajectories. At a single point, \citet{drozdz2017amplitude} linked the amplitude modulation to the convection velocity of the small scales.

Here we propose to obtain the multiscale celerity from an image couple by applying an isotropic log-Gabor filter \citep{field1987relations} at several wavelengths in the Fourier domain. 
Each scale is characterized by its wavelength: the decomposition is a measurement device, and the celerity of a band describes the footprint at that scale.

We know of no method that measures several bands on the same wall points and keeps the celerity of each band as a local quantity, which can then be correlated with the flow above it. What is new here is the combination: the features are found once, on the broadband, and each band is measured around the same points, so that position, scale and time stay in correspondence along a track. This paper focuses on the measurement. Two companion papers use its contents: one on the wall stress and pressure of a turbulent channel, to show why the multiscale celerity can be considered as a variable of the flow \citep{miozzi2026part1}; the other on a TSP dataset below the separated flow of a backward-facing ramp \citep{miozzi2026part2}. 

The paper is organized as follows:
\S~\ref{sec:datasets} introduces the datasets used in the paper for benchmarks and tests. \S~\ref{sec:class} describes the class of trackers and what they share, and in \S\ref{sec:phase} the phase formulation, the newer of the two. \S~\ref{sec:known} tests every choice of both on \emph{ad hoc} benchmarks, and compares them with PIVlab \citep{thielicke2014}, a reference PIV correlation-peak code (table~\ref{tab:pivlab}). How noise enters the displacement, and what interval of the pair it requires, is treated in \S\ref{sec:eq-noise} and \S\ref{sec:bench-noise}. \S~\ref{sec:known} closes with the stress and the pressure in their own evolution from a DNS turbulent channel, where the two tracker's formulations and the $k$--$\omega$ spectrum are compared (\S\ref{sec:dns}). \S~\ref{sec:tsp} applies the trackers to a thermal footprint in water upstream of a separating ramp, and shows how the noise of the processing chain sets the interval of the pair; \S~\ref{sec:equation} compares them with the WTH, and explores why two estimators that solve the same equation can return different celerities. \S~\ref{sec:design} collects the rules for the design of a measurement. \S~\ref{sec:conclusions} concludes. Appendix~\ref{app:coating} reports about the thermal response of the coating.
\begin{figure}
	\centering
	\includegraphics[width=.8\textwidth]{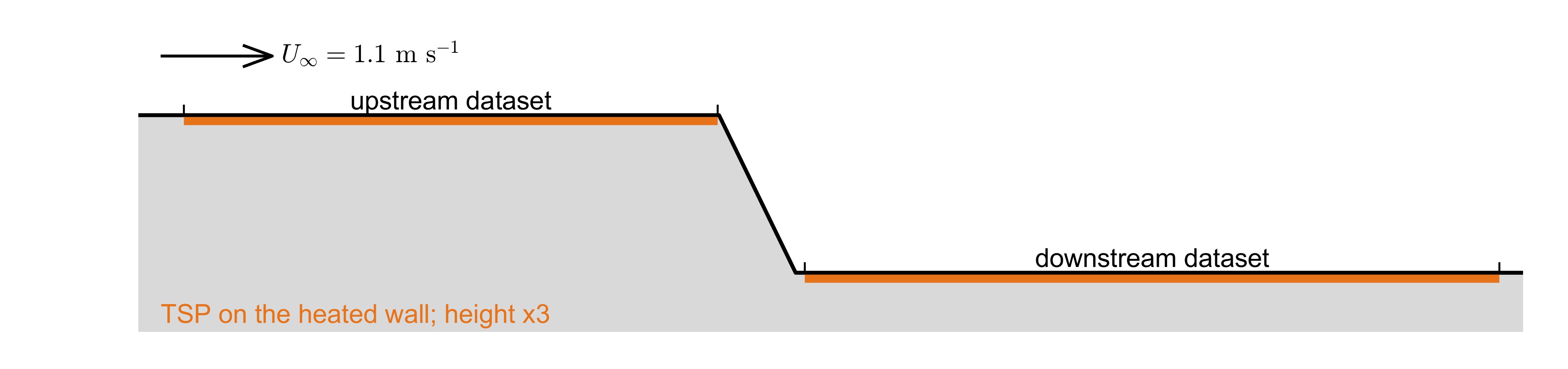}
	\caption{The dataset of \S\ref{sec:record}. The backward-facing ramp, to scale along the wall, with the height exaggerated three times: the shoulder at $x = 0$, the ramp of $25^\circ$ and $8$~mm, the two fields of view under the temperature-sensitive paint on the heated wall, and the wavelengths of the three bands.
	}
	\label{fig:ramp}
\end{figure}
\section{The datasets}\label{sec:datasets}
\subsection{The DNS stress and pressure fields}\label{sec:fields}
The data prepared for the benchmarks, and the comparisons of \S\S\ref{sec:dns} and \ref{sec:wt-compare},
use the turbulent channel of the Johns Hopkins Turbulence Database \citep{graham2016web}. That turbulent channel is a Direct
Numerical Simulation at $\Ret = 999.35$ ($1000$ in the following), driven by a constant mean pressure gradient $dP/dx=0.0025$. It is periodic along the flow and across it over $8\pi\delta \times 3\pi\delta$, with $\delta$ the half-height, on a grid of $2048 \times 512 \times 1536$ points, $12.3\,\lp$ along the flow, $6.1\,\lp$ across it and from $0.017\,\lp$ at the wall to $6.2\,\lp$ at the centreline;
its fields are stored every $0.32\,t^+$, over about one flow-through time.

The wall footprint of the streamwise velocity $u$ comes from $y^+ = 0.2193$, where its mean profile is linear, and the streamwise velocity is proportional to the wall stress,
\begin{equation}
	\tau_w/\rho \;=\; \nu\,\partial u/\partial y \;\simeq\; \nu\,u/y ,
	\label{eq:tau}
\end{equation}
so the tracked pattern is that of the stress.  

The pressure wall footprint comes from the same level at $y^+ = 0.2193$. There, the pressure equals the wall pressure within $0.2\%$ of its root mean square. The pressure is defined up to a constant at each instant, so the mean of each frame is removed.

The velocity and pressure snapshots are brought from the moving grid of the stored data back to the wall frame by a Fourier translation. The spanwise direction is decimated by two to give a square pixel of $12.26\,\lp$, and frames are taken every fourth snapshot, $\delta t^+ = 1.30$, for a displacement of about one pixel. The base analysis of stress and pressure uses a block of $500$ consecutive snapshots. The trackers run on $125$ frames (one every four snapshots), $124$ pairs and $4.8\times10^6$ tracked features. About $3.9\times10^4$ features per pair carry a celerity field on every band; the $k_x$~--~$\omega$ spectrum uses all $500$ snapshots.
For the wall pressure the $k_x$--$\omega$ spectrum uses a longer block of $1000$ frames at the same
$\delta t^+ = 1.30$, for a total duration $T^+ \approx 1300$: its coarse bands travel at up to $18\,\utau$ on $\lambda_x^+ \approx 1300$,
and the frequency bin must stay small against their frequency (\S\ref{sec:design}). The trackers use the $124$ pairs
of the first block for both fields. The stress is tracked on five bands, $\lambda_s = 4$ to $64$~px, and the pressure on
six: the pressure has more energy at the largest scales, and the band at $128$~px is added to reach the logarithmic
layer \citep{miozzi2026part1}.

The celerities of the wall pressure and of the wall stress are extracted on the same flow, at the same points and instants, with the same trackers. They are given in units of the friction velocity, $\CU = U/\utau$ for the stress and $C_p = U/\utau$ for the pressure; $C$ stands for either. 

\subsection{The temperature field}\label{sec:record}
The temperature footprint is that of \citet{costantini2026ramp}. A turbulent boundary
layer in water, at $U_\infty = 1.1$~m\,s$^{-1}$, separates over a
backward-facing ramp of $25^\circ$ and $8$~mm height and reattaches
downstream. The wall is heated at constant flux beneath a
temperature-sensitive paint and imaged at $1$~kHz. The flux of $3$~kW\,m$^{-2}$ raises the wall by about $\Delta T = 1$~K
above the water, and the Richardson number $\mathrm{Ri} = g\,\beta\,\Delta T\,h_r/U_\infty^2$, with $\beta$ the thermal
expansion coefficient of the water and $h_r$ the height of the ramp, is about $10^{-5}$: the temperature is a passive scalar.
The boundary layer begins $1000$~mm upstream of the ramp shoulder and the
heating at $x = -212$~mm. Two datasets of $12.5$~s were
recorded, upstream of the shoulder, $x = -120$ to $0$~mm, and downstream, $19$ to
$175$~mm, on a grid of $0.58$~mm; the conditions are collected in
\citet[][\S II]{miozzi2026part2}. Only the upstream dataset is measured here: at the reference position the boundary
layer is attached and the pressure gradient nearly zero. The downstream one appears only in figure~\ref{fig:tracking}cf,
and the separated flow is the subject of \citet{miozzi2026part2}. Three bands are tracked, $\lambda_s = 8$,
$16$ and $32$~px, that is $4.6$, $9.3$ and $18.6$~mm. The celerity is given as $\CT = U/\utau$, with $\utau = 47.9$~mm\,s$^{-1}$ at the reference position, $x = -97.5$~mm
\citep{miozzi2026part2}. Figure~\ref{fig:ramp} sketches the ramp and the two datasets.

\section{A class of Lagrangian trackers}\label{sec:class}

\subsection{Seeds, solvers and the scale hierarchy}

The trackers, the intensity and phase formulations of YATTS, follow
features from one frame to the next, firstly at broadband and then band by band. In agreement with the classical approach, the tracked features
are the corners of \citet{shi1994}, detected on the first raw frame. They are the points
where the normal matrix of the Lucas--Kanade minimisation is well conditioned, so
that the minimisation is unique and stable there. The same features serve every
scale and both formulations, so no difference between scales or
formulations comes from the features detector. On a band-passed field from experimental images, the
guarantee of well conditioning becomes a presumption, and the fraction of features on which the
minimisation converges, the yield, measures how well it holds.

The intensity solver is the affine Lucas--Kanade one \citep{lk1981,baker2004}. The method assumes that a feature in frame $A$ can be tracked to frame $B$ by minimizing the SSD within a window of radius $R$ centred on the feature in both frames. The minimization is performed under a linear warp using a Gauss-Newton iterative scheme and a forward-backward consistency check \citep{kalal2010forward}. Each feature gives one displacement per frame pair and per scale, and the whole process returns the instantaneous celerity at each feature.
Together, the celerities form the sparse field $C(x,z,t;\lambda)$ and, over the dataset, a population that can be described statistically, from which a median, a robust scatter and a yield are taken. 
\begin{figure}
	\centering
	\includegraphics[width=.9\textwidth]{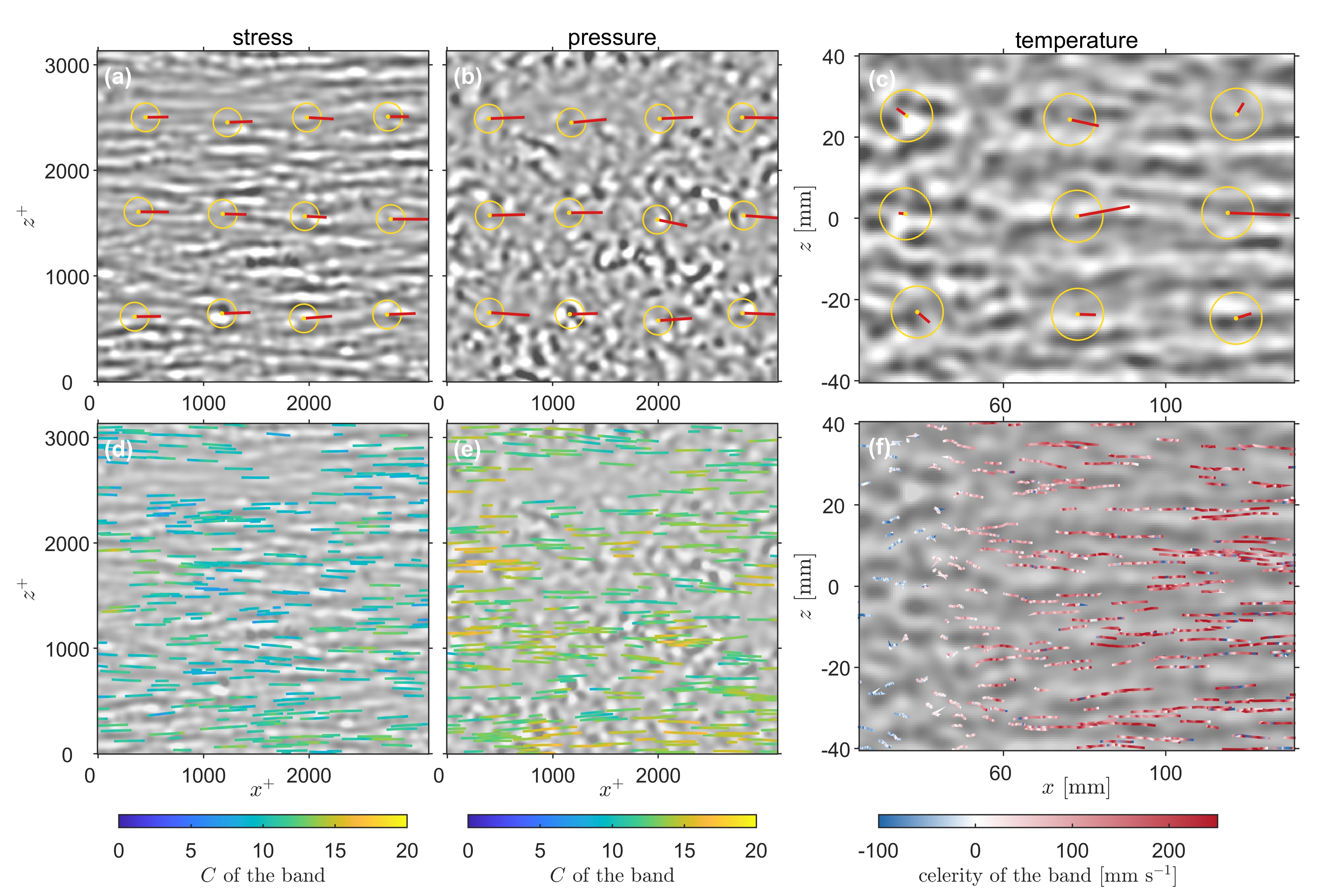}
	\caption{The same tracker on three footprints, on the middle band, $\lambda_s = 16$~px. Columns: the wall stress and the wall pressure of the turbulent channel, at the same instant and on the same square of $3140\,\ell^+$, and the thermal footprint of the dataset downstream of the ramp (\S\ref{sec:record}). Top row: the band in gray; twelve features, each with the window of the band (circle, $R = 11$~px) and the displacement of the band over one pair, magnified $20$ times on the channel and $8$ times on the TSP. The affine deformation over one pair is about one per cent and is not visible. Bottom row: $300$ tracks of features, drawn at random among those of at least five pairs, over $20$ pairs ($26\,t^+$ on the channel, $120$~ms on the TSP). They are coloured by the celerity of the band, in friction velocities on the channel and in mm\,s$^{-1}$ on the TSP. On the TSP the tracks upstream of $x \approx 37$~mm point upstream, inside the recirculation.
	}
	\label{fig:tracking}
\end{figure}
We detect the features on the raw frame, the broadband. Tracking is split in two steps: firstly, features are carried by the intensity solver under translational warp. Then, the final displacements are refined using the affine warp under the intensity or phase constancy constraint. At each time, the new points on the trajectories are therefore built on the broadband. After solving the broadband, every band is measured on the same features, each with its own window, starting from the broadband refined displacement as initial seed, using the affine intensity or phase constancy constraint. The displacement of a band gives its celerity, $C(x,z,t;\lambda)$, but is not used to move the points, and at the next pair the window of each band is centered again on the broadband position. As a final outcome, every point of a track knows the celerity of all the bands at that point, while the track itself belongs to the trajectory evaluated at broadband scale. The celerity of a band is thus the displacement of that band over one pair, at points carried by the broadband; it is not the velocity of a trajectory of the band. There are no reasons that mandate this description of the trajectories. As an already available alternative, the features identification could follow a per--band approach. In this way, each band would provide its own set of features and trajectories, but for the present work we chose to drive the trajectories' description from the broadband because of the possibility of a direct comparison among different scales, its simplicity, and the efficiency in managing the dataset. The benchmarks and the turbulent channel rest on single pairs, so they measure the accuracy of the displacement, whatever the trajectory; the trajectory sets only where a band is measured. The track median of \S\ref{sec:tsp} follows the broadband.

Figure~\ref{fig:tracking} shows the windows and the tracks on the three footprints used here, for one band. On the turbulent channel DNS (\S\ref{sec:fields}) every result rests on single pairs; on the TSP dataset of the backward facing ramp (\S\ref{sec:record}) each band's celerity is filtered along the tracks of the features
(\S\ref{sec:tsp}).

A video of the tracking results at several bands, for the wall footprint of stress and pressure, is available in the supplementary material (Movie 1).

\begin{figure}
	\centering
	\includegraphics[width=.6\textwidth]{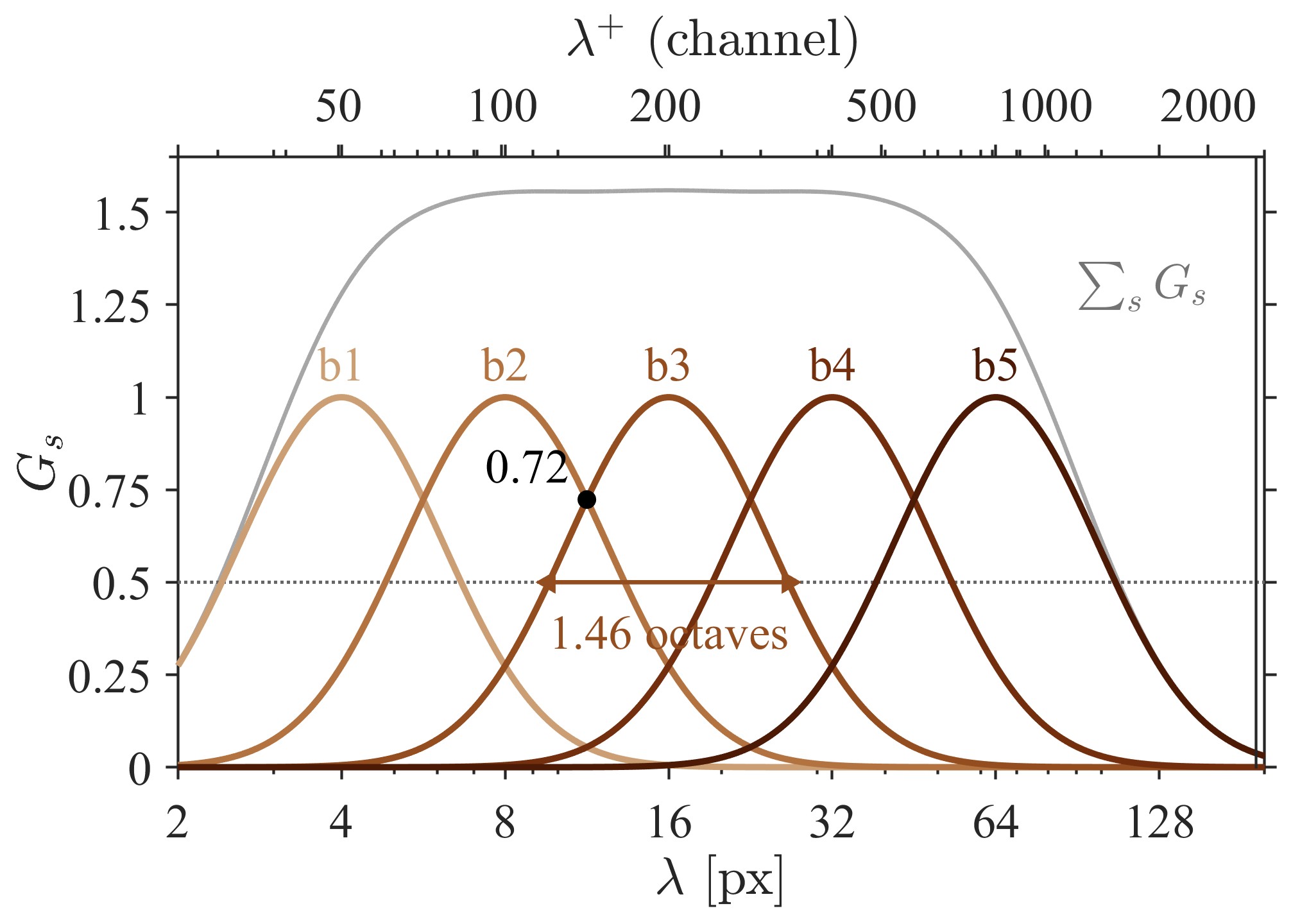}
	\caption{The isotropic log-Gabor filter in equation~\eqref{eq:loggabor} of the five bands common to the stress and the pressure, along any radius of the wavenumber plane, against the wavelength $\lambda = 2\pi/|\mathbf{k}|$; top axis, the same wavelength in wall units of the channel. Each band is $1.46$ octaves wide at half maximum, from $\lambda_s/1.66$ to $1.66\,\lambda_s$ (arrow, b3), and bands one octave apart cross at $0.72$ of their peak (dot). Grey: the sum of the five filters, constant within half a per cent between the peaks of b2 and b4. One octave is a factor of two in wavelength.
	}
	\label{fig:loggabor}
\end{figure}

Scales are separated by an isotropic log-Gabor filter in the Fourier
domain \citep{field1987relations},
\begin{equation}
	G_s(\mathbf{k}) \;=\;
	\exp\!\left[-\frac{\ln^2\!\big(|\mathbf{k}|\,\lambda_s/2\pi\big)}{2\ln^2\sigma}\right],
	\qquad \sigma = 0.65,
	\label{eq:loggabor}
\end{equation}
which peaks at $|\mathbf{k}| = 2\pi/\lambda_s$, has no zero-wavenumber term,
and its full width at half maximum is $2\sqrt{2\ln 2}\,|\ln\sigma|/\ln 2 \approx 1.46$ octaves: $G_s$ falls to one half at $\lambda_s/1.66$ and $1.66\,\lambda_s$ (figure~\ref{fig:loggabor}). 

Bands one octave apart cross at $0.72$ of their peak, and the sum of the bands is flat between b2 and b4, so the decomposition favours no inner scale. The filter is chosen for three
properties: i) It is linear, so a pattern that moves rigidly moves rigidly in
every band. ii) Its band is fixed in wavelength, so a celerity belongs to a
scale, and the same scale can be compared with another flow. iii) And its radial
profile builds the quadrature bank of the phase formulation
(\S\ref{sec:constraints}).
Five octave-spaced bands are used
on the turbulent channel, and three on the backward facing ramp of \citet{miozzi2026part2}.
Each band is measured on its own window, so the locality of an estimate is
set by its band.

The tracker is Lagrangian with respect to the footprint: a feature is a place in a pattern
and does not follow a material point of the fluid, and its celerity is the speed of the pattern. On a dispersive field it is the phase
celerity of the band, not the group velocity of a packet
(\S\ref{sec:mcdecay}).

\subsection{The Brightness Constancy Constraint and the intensity tracker}\label{sec:constraints}
The \emph{intensity} formulation aims to find a solution preserving the brightness of the interrogation window from image $A$ to image $B$ \citep{lk1981}. The band-passed image is real, and the isotropic band introduces no
directional bias (figure~\ref{fig:bank}a,d). Its window radius is free, and the phase formulation uses the same. Unlike in the PIV approach, it cannot be related to a density of the particles falling within the window. It must be large enough for the minimisation to converge, which governs the two finest bands, and small enough not to blur a varying celerity, which governs the coarse ones. We chose $R = 6$, $8$, $11$, $21$ and $42$~px after a sweep on the benches (\S\ref{sec:radius}). On the two finest bands the radius is the smallest at which the number of converged features stops growing; on the three coarser bands it is $0.65\,\lambda_s$ ($84$~px on the sixth band of the pressure).

\subsubsection{The cost and its normal equations}\label{sec:eq-lk}
The tracker asks for the warp $\mathbf{W}(\bm{x};\mathbf{p})$ that best maps one image onto the next, on a window $W$ around a feature, between two frames separated by a time $\tau$. Its cost is the sum of squared differences,
\begin{equation}
	E(\mathbf{p}) = \sum_{W} w(\bm{x}) \left[I_1\big(\mathbf{W}(\bm{x};\mathbf{p})\big) - I_0(\bm{x})\right]^2 ,
	\label{eq:ssd}
\end{equation}
with $w$ the window weight. The warp is affine, $\mathbf{W}(\bm{x};\mathbf{p}) = \bm{x} + \bm{d} + \mathbf{L}\,\bm{x}$, with $\bm{x}$ measured from the feature and six parameters $\mathbf{p}$: the two components of the displacement $\bm{d}$ and the four displacement gradients of $\mathbf{L}$. Linearised about the current estimate, the minimum obeys
\begin{equation}
	\mathbf{G}\,\delta\mathbf{p} = \bm{e} ,\qquad
	\mathbf{G} = \sum_W w\,\mathbf{J}^{\mathsf T}\mathbf{J} ,\qquad
	\bm{e} = \sum_W w\,\mathbf{J}^{\mathsf T}\left[I_0 - I_1(\mathbf{W})\right] ,\qquad
	\mathbf{J} = \nabla I_0^{\mathsf T}\,\frac{\partial\mathbf{W}}{\partial\mathbf{p}} ,
	\label{eq:lk}
\end{equation}
with the Jacobian taken on the first image, so that $\mathbf{G}$ is computed once per level, and the warp updated by
composition, $\mathbf{W} \leftarrow \mathbf{W}\circ\mathbf{W}(\,\cdot\,;\delta\mathbf{p})$: the inverse compositional
scheme of \citet{baker2004}, with the increment inverted to first order \citep{miozzi2008}.

For a translation, $\mathbf{L} = 0$ and $\partial\mathbf{W}/\partial\bm{d}$ is the identity, so that $G_{ij} = \sum_W w\,\partial_i I_0\,\partial_j I_0$ and $e_i = \sum_W w\,\partial_i I_0\,(I_0 - I_1)$; this is the form used in \S\ref{sec:eq-noise} and \S\ref{sec:eq-weak}.

The warp is fitted in two stages. The translation is iterated from the coarsest level of a pyramid down to full resolution, with a forward--backward check \citep{kalal2010forward}. The affine warp then refines it at full resolution only, starting from the translation; where it does not converge, the translation is kept. The bands use no pyramid: each starts from the seeds provided by the broadband displacement at full resolution.

\subsubsection{Noise and the interval of the pair}\label{sec:eq-noise}
Noise matters to the celerity estimation because the displacement over one time step is small. On the datasets of this paper it is one to two pixels, so a bias of a tenth of a pixel is a bias of five to ten per cent on the celerity. Images based on fluorescence are noisy. We take the noise as additive, with zero mean, independent of the signal and stationary over the window and the interval; before the filters it is white. Saturation and errors of registration are outside this model. Images are cleaned before any estimate: the chain of the TSP dataset applies a spatial and a temporal filter (\S\ref{sec:tsp}). Two questions follow. Does the noise bias the tracker, or only spread its estimates? And does the cleaning attempt change the response of the trackers? To answer the questions, each image is split into signal and noise, to inspect where the noise enters the solver, and where it moves the minimum that the iteration converges to. Under these assumptions, noise independent between the frames only spreads the estimates, while noise correlated between the frames shifts them. A temporal filter correlates the noise of frames closer than its length, even when the noise is white before its application. To avoid this bias, the time interval separating the frames must exceed the filter length. The benches test this bound (\S\ref{sec:bench-noise}) and the processing of the TSP dataset is build around it (\S\ref{sec:tsp}).

Write each image as signal plus noise, $I = S + n$, with $n$ of variance
$\sigma^2$ and zero mean.
\begin{figure}
	\centering
	\includegraphics[width=.95\textwidth]{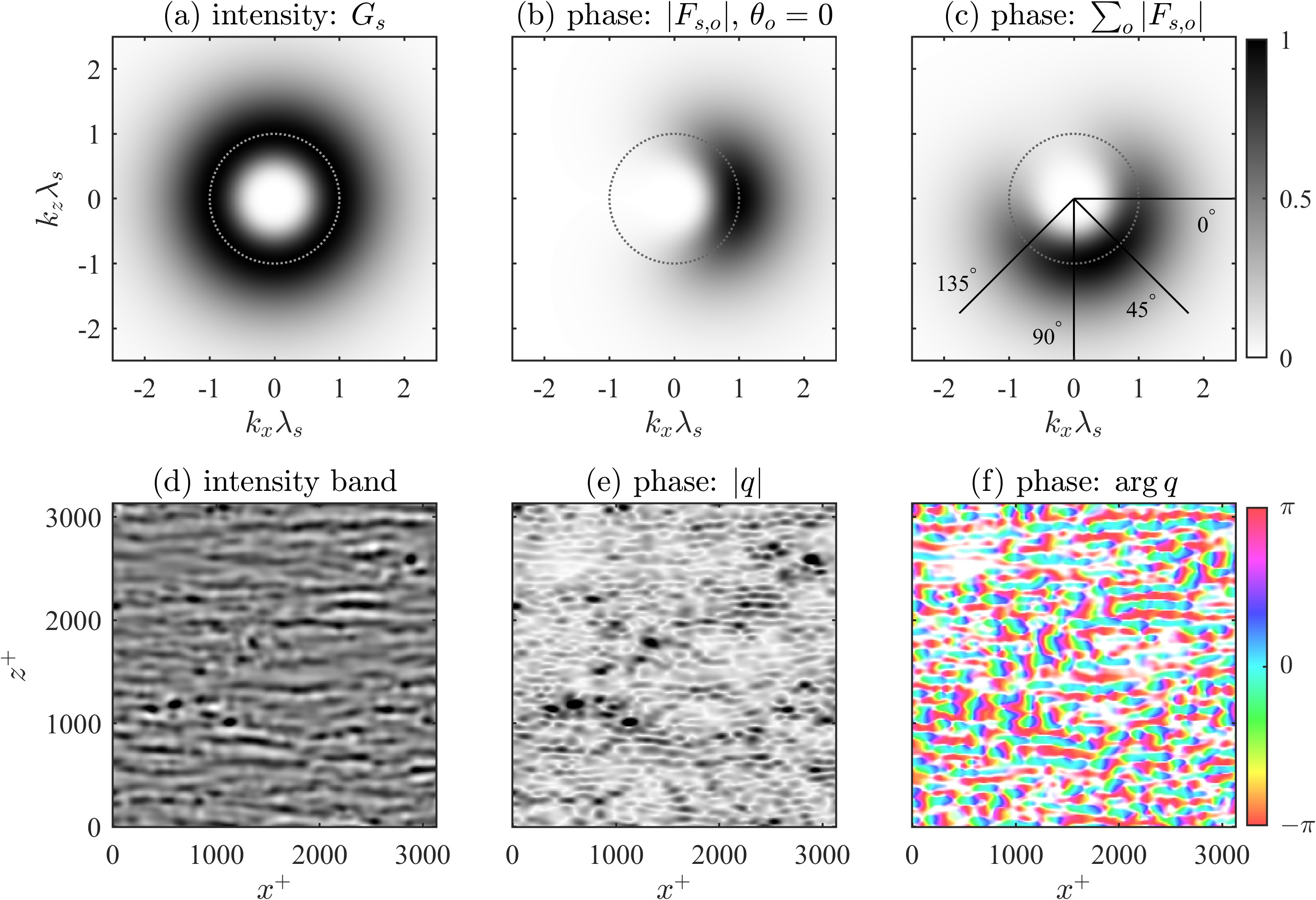}
	\caption{The two formulations on one band, $\lambda_s^+ = 196$. Top row, the
		filters in the wavenumber plane, normalised by the band wavenumber (dotted
		ring): (a) the isotropic radial log-Gabor $G_s$ of the intensity tracker;
		(b) one channel of the phase tracker, $\theta_o = 0$; (c) the four channels
		summed, with the channel directions drawn: the lobes are one-sided, so
		together they cover a half-plane, which is all of it for a real field, and it
		is the one-sidedness that gives the response in (f) a phase whose
		gradient is the local wavevector. Bottom row, a $3140\,\ell^+$ square of the turbulent channel plane at
		$y^+ = 0.22$: (d) the snapshot through (a), the quantity held constant by
		the intensity tracker; (e) amplitude and (f) phase of the snapshot through
		(b), (f) faded where the amplitude (e) is weak: the phase tracker
		holds (f) constant and is invariant to (e).
	}
	\label{fig:bank}
\end{figure}
In the tracker the noise in $\mathbf{G}$ does not bias the result. $\mathbf{G}$ sets the length of each step but not the point where the iteration stops: at convergence $\bm{e} = 0$, the minimum of \eqref{eq:ssd}, and the result depends on where that minimum lies. Its expected value is
\begin{equation}
	\langle E(\bm{d})\rangle = \sum_W w \left[S_1(\bm{x}+\bm{d}) - S_0(\bm{x})\right]^2
	+ 2\sigma^2 \sum_W w
	- 2\,C_n(\bm{d},\tau) \sum_W w ,
	\label{eq:ssd-mean}
\end{equation}
where $C_n(\bm{r},\tau) = \langle n_1(\bm{x}+\bm{r})\,n_0(\bm{x})\rangle$ is the
covariance of the noise between the two frames. The cross term between signal and noise has zero mean and spreads the result without shifting it; the second term is a constant, while the third depends on $\bm{d}$.

If the noise of the two frames is independent, $C_n = 0$ and the minimum of
\eqref{eq:ssd-mean} is the minimum of the signal: the tracker is unbiased
however noisy the images are. But a temporal filtering before the tracking removes this independence. Let the noise be white, and let it
pass through a temporal filter of taps $h_j$ and a spatial filter. After both, it has variance $\sigma^2$, normalised
spatial correlation $K(\bm{r})$, and covariance between two frames
\begin{equation}
	C_n(\bm{r},\tau) = \sigma^2 \rho(\tau)\,K(\bm{r}) ,\qquad
	\rho(\tau) = \frac{\sum_j h_j h_{j+\tau}}{\sum_j h_j^2} .
	\label{eq:cn}
\end{equation}
The cost \eqref{eq:ssd}, the weighted sum over the window of the squared differences between the first frame at $\bm{x}$
and the second at $\bm{x}+\bm{d}$, contains the product $n_1(\bm{x}+\bm{d})\,n_0(\bm{x})$; its mean
\eqref{eq:ssd-mean} therefore takes this covariance at $\bm{r} = \bm{d}$.

The noise does not move with the pattern, so its term in equation \eqref{eq:ssd-mean} is centered at $\bm{d} = 0$: a ridge for
$\rho(\tau) < 0$, which pushes the minimum along the motion, and a trough for
$\rho(\tau) > 0$, which pulls it toward zero. Near the true displacement $d_0$, along $x$, the signal term is the
parabola $a\,(d-d_0)^2$, with $a = \sum_W w\,(\partial_x S)^2$. Setting to zero the derivative of the sum of the two
terms, the minimum moves by
\begin{equation}
	\delta \simeq \frac{\sigma^2 \rho(\tau)\,K'(d_0)\sum_W w}{a}
	= \rho(\tau)\,\frac{N}{A_s}\,\frac{K'(d_0)}{-K''(0)} ,
	\label{eq:bias}
\end{equation}
with $N = \sigma^2\left(-K''(0)\right)\sum_W w$ the gradient energy of the noise
in the window and $A_s = a$ that of the signal. Since $K' < 0$ for displacements below half a wavelength, the shift has the sign opposite
to $\rho$. It grows with the share of noise $N/A_s$, which is large in the streamwise direction, where the gradients
of a streaky footprint are weak. For $d_0 \ll \ell$, with $\ell$ the width of $K$, the relative error tends to
$\delta/d_0 = -\rho\,N/A_s$; for $d_0 \gg \ell$ it vanishes.

The interval in time is the key mechanism to overcome the issue. A filter of $M$ taps correlates the noise only up to the lag $M-1$, so $\rho(\tau) = 0$ for $\tau \ge M$: the interval of the pair must exceed this reach, or fall where $\rho$ is negligible: with the seven taps of the TSP dataset
$|\rho| < 0.003$ from five frames on, and the pairs are six frames apart (\S\ref{sec:tsp}). Adding the known term back to the cost would allow shorter lags, at the price of an estimate of $\sigma^2$; it was not needed here.

The phase formulation meets the same term: its residual is the argument of
$q_2 q_1^{*}$, whose noise part has the mean $\sigma^2\rho(\tau)K$ of equation
\eqref{eq:cn}, so the minimum moves with the same sign, weighted by the
amplitudes.
The turbulent channel tests these predictions with the intensity tracker in \S\,\ref{sec:bench-noise}.
\subsection{The Phase Constancy Constraint and the phase tracker}\label{sec:phase}
The \emph{phase} formulation holds constant the phase of a log-Gabor
quadrature filter bank \citep{fleet1990}, with the same radial profile in four
one-sided orientation channels (figure~\ref{fig:bank}b,c). We are interested in this approach because the phase
does not depend on a uniform change in the amplitude of the scalar. The gradient of
the phase is the local wavevector, normal to the lines of constant phase, so one channel at one point gives only the
component of the displacement along it. The same holds for the intensity, where one point gives only the component
along the intensity gradient. The intensity tracker recovers the full displacement from the different directions its
gradient takes within the window; the phase tracker sums over the window and over the four channels, which look at the
pattern along four directions at every point (\S\ref{sec:phasec}). Its window is that of the intensity tracker, band by band
(\S\ref{sec:constraints}).

\subsubsection{The quadrature bank}\label{sec:qbank}
Each channel $o$ of the bank multiplies the radial profile \eqref{eq:loggabor}
by an angular Gaussian centered on the direction $\theta_o$,
\begin{equation}
	Q_{s,o}(\mathbf{k}) \;=\; 2\,G_s(\mathbf{k})\,
	\exp\!\left[-\frac{(\theta_{\mathbf{k}}-\theta_o)^2}{2\,(\Delta\theta\,\pi/O)^2}\right],
	\qquad \theta_o = o\,\pi/O,\quad o = 0,\dots,O-1,
	\label{eq:bank}
\end{equation}
with $\theta_{\mathbf{k}}$ the direction of $\mathbf{k}$, $O = 4$ channels and
$\Delta\theta = 1.2$. The Gaussian is not mirrored to $\theta_o + \pi$: the
filter is one-sided, and the response of a real image,
$q_{s,o} = \mathcal{F}^{-1}\big[\hat I\,Q_{s,o}\big]$, is complex, an analytic
signal along $\theta_o$. Its modulus $A = |q|$ is the local amplitude and its
argument $\phi = \arg q$ the local phase. The four directions cover half the
wavenumber plane, which for a real field is all of it (figure~\ref{fig:bank}c).
The angular width is chosen large: neighbouring channels share modes, and
no direction of the pattern rests on one channel alone.

\subsubsection{The Phase Constancy Constraint}\label{sec:phasec}
The tracker holds the local phase of every channel of a band constant along the warp,
\begin{equation}
	\phi_{o}\big(\mathbf{W}(\mathbf{x};\mathbf{p}),\,t_2\big) = \phi_{o}(\mathbf{x},\,t_1) ,
	\label{eq:phasec}
\end{equation}
with $\mathbf{W}$ the affine warp of the intensity tracker, whose six
parameters $\mathbf{p}$ include the displacement. Neither the residual nor the
spatial gradient of the phase needs unwrapping:
\begin{equation}
	r_o = \arg\!\big[q_o(\mathbf{W},t_2)\,q_o^{*}(\mathbf{x},t_1)\big] ,\qquad
	\nabla\phi_o = \frac{\mathrm{Im}\big(q_o^{*}\,\nabla q_o\big)}{|q_o|^2} .
	\label{eq:phasegrad}
\end{equation}
The gradient is the local wavevector of the channel. Linearised about the
current warp, equation \eqref{eq:phasec} gives one equation per pixel and channel, with
the gradient of the phase where equation \eqref{eq:lk} has that of the intensity. The
normal equations sum over the window and over the channels,
\begin{equation}
	\mathbf{G}\,\delta\mathbf{p} = \bm{e} ,\qquad
	\mathbf{G} = \sum_W \sum_o w_o\,\mathbf{J}_o^{\mathsf T}\mathbf{J}_o ,\qquad
	\bm{e} = -\sum_W \sum_o w_o\,\mathbf{J}_o^{\mathsf T} r_o ,
	\label{eq:phaselk}
\end{equation}
with $\mathbf{J}_o = \nabla\phi_o^{\mathsf T}\,\partial\mathbf{W}/\partial\mathbf{p}$
and the weight $w_o = A_o(\mathbf{x},t_1)\,A_o(\mathbf{W},t_2)$. A phase is reliable where the band has energy \citep{fleet1993}, and the product of the two amplitudes weights each constraint by that reliability in both frames. The amplitude enters only the weights. The constraint does not change when the scalar fades or grows uniformly, which is the purpose of the formulation.

\subsubsection{Iteration and rejection of pixels and channels contribution}\label{sec:phaserej}
The phase solver has two stages. A translational intensity tracker on the raw
frames, with a window of $10$~px, a pyramid of three levels and a
forward--backward check, gives a first, potentially large displacement within half the finest
wavelength and rejects the features that fail the check. Each band is then
solved on its own four channels, in its own window, using the initial seed from that
displacement, by Gauss--Newton with a light Levenberg damping, at most $150$
iterations, until the displacement at the edge of the window changes by less
than $10^{-3}$~px. On the coarser bands the window spans about $1.3$ wavelengths; on the bands at $4$ and $8$~px it spans three and two, the radii chosen for the intensity tracker (\S\ref{sec:radius}). Equation~\eqref{eq:phasec} is imposed at points at most $\lambda_s/4$ apart, the Nyquist limit of the band.

A phase is not defined where the amplitude vanishes, and near such points it
varies faster than the band allows \citep{fleet1990}. Three tests remove a
pixel and channel from the sums. Its amplitude must exceed five per cent of
the median of the window. The modulus of its phase gradient must match the
wavenumber of the band within $60$~per~cent, the singularity test of
\citet{fleet1990}. And its phase step between the frames must stay below
$2.6$~rad, short of the $\pi$ at which the argument wraps. A feature with
fewer than twelve usable constraints is dropped.
\subsection{The intensity and phase tracker formulations side by side}\label{sec:both}
Both intensity and phase tracker formulations resample the second frame with a quintic B-spline
\citep{thevenaz2000}, and work on the same band content. They differ only
in the quantity held constant, and neither uses any part of the other's
objective function. A result on which they agree does not depend on the quantity held constant. The agreement does not test what they share: the features, the broadband seed, the bands, the windows and the affine warp.

The benchmarking results indicate that the intensity tracker framework is the tool of choice, combining an execution seven times faster, a higher yield of tracked features, and improved coverage of the spatial domain near the boundaries.

\section{Benchmarks}\label{sec:known}

The trackers are tested on fields known at every point. Such a field can be generated or taken from a
real one. We started working with a Gaussian surrogate, by preparing a random field with the spectrum of the real one, but it has no skewness, while the wall stress is strongly skewed and the decay of the footprint acts mainly on
its intense, isolated events. We therefore start from real frames of the turbulent channel plane (\S\ref{sec:fields}) and
impose on them one property at a time, taking it from the list of the changes we are interested in: a uniform shift, a celerity that varies with the scale or along the wall, a decay, the response of a coating, a noise correlated in time. The first frame of each pair is a real one, with its skewed, intermittent pattern, and the second is made from it, so that the imposed celerity is known. These fields test the estimator and not the evolution of real turbulence, which is tested last, on consecutive frames of the turbulent channel (\S\ref{sec:dns}). 

\subsection{The synthetic method}\label{sec:mcmethod}

Each pair is made of a real frame of the stress footprint of the channel, with its mean removed, $A$, and of its image $B$ obtained by applying four kinds of degradation, evaluated after one interval $\Delta t$. The consequences on the celerity are tested one at a time: the dispersion alone, each of the two decays together with a uniform shift of $1.071$~px ($p = 0$), and the warp alone. The decay linked to the stress is applied after the shift.

\emph{Dispersion and uniform decay.} They are imposed on the Fourier transform $\hat A$ of the frame,
\begin{equation}
  \hat B(\mathbf{k}) =
  \hat A(\mathbf{k})\,
  \mathrm{e}^{-\mathrm{i}\omega(\mathbf{k})\,\Delta t}\,
  \mathrm{e}^{-\Delta t/T_d(|\mathbf{k}|)},
  \qquad
  \omega = c(|\mathbf{k}|)\,k_x, \qquad
  T_d \propto |\mathbf{k}|^{-2}.
  \label{eq:operator}
\end{equation}
The first exponential moves each mode at a known phase speed $c(|\mathbf{k}|)$ along $x$; the second reduces its
amplitude, by a factor $\mathrm{e}$ in the time $T_d(|\mathbf{k}|)$. The speed is a power law,
$c = c_0\,(|\mathbf{k}|/k_0)^{-p}$, with $c_0\,\Delta t = 1.071$~px at $2\pi/k_0 = 16$~px. The contrast between the
coarsest band, $\lambda = 64$~px, and the finest, $\lambda = 4$~px, is $c_5/c_1 = 16^{\,p}$; $p = 0$ is no
dispersion. The rate $1/T_d$ grows as $|\mathbf{k}|^2$, as in diffusion, and is the same at every point. It is fixed by
the fraction of the amplitude that the finest band retains after $\Delta t$, $0.85$ or $0.70$.

\emph{Decay linked to the stress.} In a thermal footprint the heat transfer grows with the wall stress, and the pattern
fades faster where the stress is high. Such a decay depends on the position, so it is imposed in physical space: $B$ is
multiplied, at each point, by
\begin{equation}
  \rho_C^{\,m}, \qquad m = \max\left(0,\; 1 + 0.42\,\hat a\right),
  \label{eq:decay}
\end{equation}
where $\hat a$ is $B$ divided by its rms, before this multiplication, and $0.42$ is the ratio of the rms of the stress
to its mean, so that $m$ is the local stress over its mean. Where the stress equals its mean the amplitude is reduced by
the factor $\rho_C \le 1$; where it is twice the mean, by $\rho_C^2$; where the flow reverses, $m = 0$ and the amplitude
does not change. The mean of $B$ is then removed.

\emph{Uniform shift.} All the modes move by the same displacement, through a Fourier shift.

\emph{Warp.} The displacement depends on the position, $B(x) = A\big(x - d(x)\big)$, with $d(x)$ defined at the point of
arrival.

Each of these benches is repeated on three frames of the channel (\S\ref{sec:fields}), $45\,t^+$ apart, and with both trackers on
the same features. The response of a coating and the noise correlated in time are imposed in a different way, described in
\S\ref{sec:mccoating} and \S\ref{sec:bench-noise}.

\subsection{The window radius of the intensity tracker}\label{sec:radius}

The window radius was swept on three configurations: no dispersion or decay, the warp
at a period of $512$~px, and the decay of equation \eqref{eq:decay} with $\rho_C = 0.8$. A small window loses features on the two finest bands and a large one blurs the
modulation on the coarse bands; on a uniform celerity the radius never costs
accuracy, within $0.04\%$. The choice is $R = 6$, $8$, $11$, $21$ and $42$~px
on the five channel bands, and $8$, $11$ and $21$~px on the three ramp bands
of \citet{miozzi2026part2}; the coarse bands take $0.65\,\lambda_s$.

\subsection{Response in \texorpdfstring{$k$--$\omega$}{k--omega} under decay}\label{sec:mcdecay}

The operator in equation \eqref{eq:operator} imposes $\omega(k)$, so the test of a
tracker in the $k$--$\omega$ plane is whether it returns $c(k)$ band by
band. Table~\ref{tab:mc_kw} gives the recovered contrast between the
coarsest and finest bands.

\begin{table}
\centering
\begin{tabular}{lcccc}
\toprule
case & imposed $c_5/c_1$ & intensity tracker & phase tracker & unfiltered field [px] \\
\midrule
non-dispersive & 1.000 & 1.000 & 1.000 & 1.071 \\
mild           & 1.300 & 1.243 & 1.245 & 1.000 \\
strong         & 2.000 & 1.779 & 1.801 & 0.900 \\
\bottomrule
\end{tabular}
\caption{Scale resolution: recovery of an imposed contrast between
$\lambda = 64$ and $4$~px, three frames. Both formulations separate the
scales. Last column: the intensity tracker on the unfiltered field ($R = 10$~px) returns one celerity, in px per frame; on
the two dispersive fields, that imposed at $\lambda \approx 8$~px.
}
\label{tab:mc_kw}
\end{table}

\begin{figure}
	\centering
	\includegraphics[width=\textwidth]{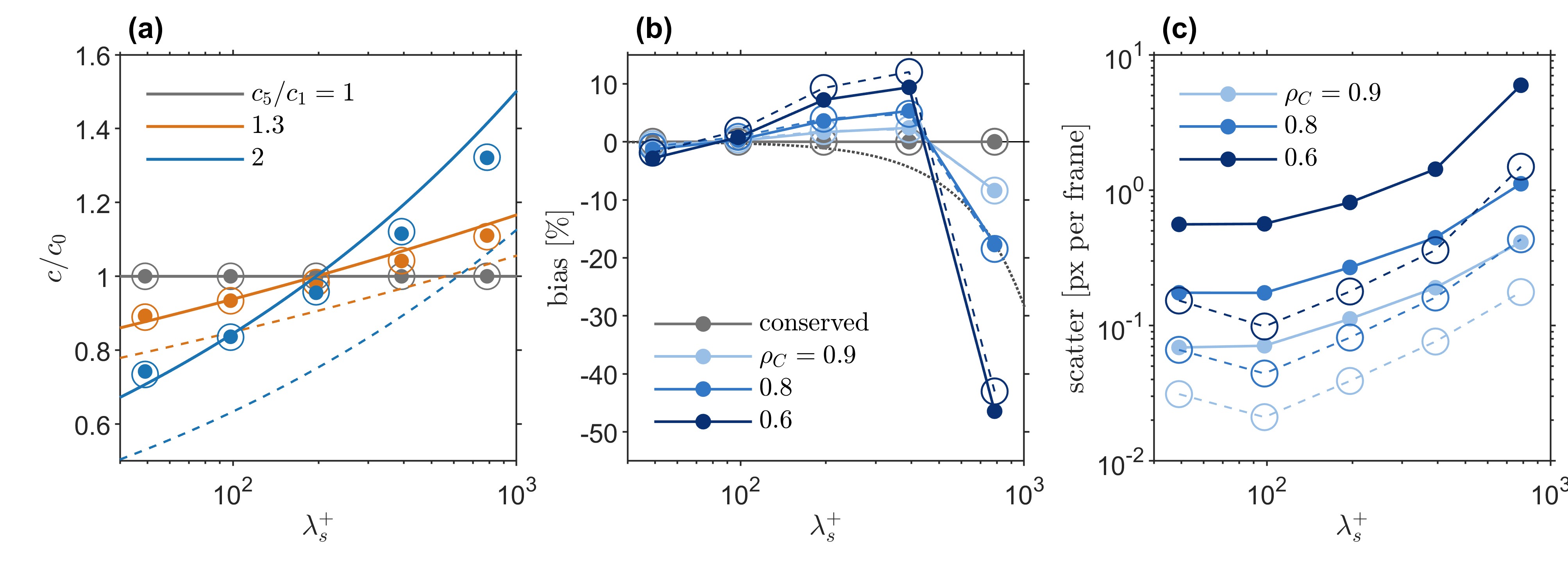}
	\caption{The response of the two trackers on the synthetic pairs of \S\ref{sec:mcmethod}, three frames, against the wavelength of the band. Filled symbols, intensity tracker; open symbols, phase tracker. (a) Dispersive fields with a contrast $c_5/c_1$ of $1$, $1.3$ and $2$ between the coarsest and the finest band: the recovered celerity over $c_0 = 1.071$~px per frame. Solid, the imposed phase celerity, $c \propto \lambda^p$; dashed, the group velocity of the modes aligned with $x$, $(1-p)\,c$. (b) Bias of the recovered celerity under the decay \eqref{eq:decay}, $\rho_C = 0.9$, $0.8$ and $0.6$, which lowers the rms to $0.86$, $0.75$ and $0.55$ of its initial value, and for a conserved field; dotted, a guide that grows as $\lambda^2$. A decay uniform in space, $T_d \propto |\mathbf{k}|^{-2}$, that removes $15$ or $30\%$ of the amplitude of the finest band, not shown, biases the intensity tracker by less than $0.2\%$ and the phase tracker by at most $1\%$, on the finest band. (c) Robust scatter under the same decays, in pixels per frame.
	}
	\label{fig:kw}
\end{figure}

On a dispersive field the phase velocity of a band and the group velocity
of a packet in it differ \citep{renard2015}, and which one a tracker on a band-passed
field returns is not obvious. With $c \propto k^{-p}$ the
group velocity of the modes aligned with $x$ is $(1-p)\,c$, a quarter below the phase velocity at the
strong contrast, and closer to $c$ for the others. It does not enter the cost \eqref{eq:ssd}, whose minimum, for a
small displacement, is the band average of $c$ weighted by the energy of the streamwise gradient. Both trackers return the imposed \emph{phase} celerity,
within $7\%$ at the strong contrast and $3\%$ at the mild one (figure~\ref{fig:kw}a). This
attenuation is the transfer function of the band, and not an error of the solver: the tracker returns the average of
$c(k)$ over the width of the filter. A narrower band reduces it. Between $\sigma = 0.55$ and $0.75$ ($2.0$ to $1.0$ octaves) the bias on a
uniform shift stays below $0.005\%$, and the recovered contrast of two goes from $1.68$ to $1.90$ with the phase
tracker and from $1.64$ to $1.89$ with the intensity one; the two trackers differ by at most $2.5\%$.

Figure~\ref{fig:kw}(b, c) gives the response under decay. A decay uniform in space
biases neither formulation by more than one per cent. The decay linked to the local stress (equation \ref{eq:decay}) biases
the coarsest band of both by about a fifth at $\rho_C = 0.8$. The cause is
the gradient of the envelope at the scale of the band, which perturbs a
phase and a band-passed intensity alike: tracking the decayed frame
against its own exact translate leaves no bias, and smoothing the envelope
removes it. The phase formulation keeps a scatter two to four and a half times lower
on the coarse bands, the more so as the decay grows.

\subsection{Uniform celerity: the noise floor}\label{sec:uniform}

A frame translated by a Fourier shift has a celerity that is exact and the
same at every point. From half a pixel to four, both trackers return
it on every band within $5\times10^{-5}$~px, at most $0.005\%$ of the shift, with a scatter below $2\times10^{-4}$~px.
 Degrading the pair to the coherence measured between real
frames raises the scatter of the broadband displacement to about $0.05$~px
and moves its median by less than a tenth of a per cent. On the bands the scatter is $0.07$ to $0.12$~px, and the median moves by less than $0.4$~per cent on b1 to b4 and by up to $1.3$~per cent on b5. The noise of real frames therefore limits the precision of the trackers well before their accuracy.

\subsection{A known non-uniform celerity: transfer and attribution}\label{sec:warp}

A celerity that varies along the wall tests two things: the share of the modulation that the tracker returns, and where
it places it. We impose the displacement $d(x) = d_0 + a\sin(2\pi x/L)$, with $d_0 = 1.071$~px, $a = 0.3$~px and
$L = 256$ or $512$~px, and project the recovered displacement, mean removed, on the same period. Both trackers return the
amplitude $a$ within six per cent, except on the coarsest band at $L = 256$~px, where the window spans a third of the
period and both lose a sixth. The position of the maximum along $x$ is found within two thousandths of the period on
average, with a scatter of the same order between the three frames, and the two trackers agree within $0.1$~px of position.
The same holds for a shear, a rotation and a strain, where $d$ varies along $z$, or along $x$ and $z$, with the gain
measured on both components of the displacement. The affine gradient of the window is $0.8$ to $1.0$ of the imposed one. These
fields do not deform as turbulence does; that case is \S\ref{sec:dns}, where no exact answer exists.

\subsection{A thermal filter due to the response of the wall coating}\label{sec:mccoating}

A paint on a heated wall is a low-pass filter: it lets the slow changes of temperature through and damps the fast ones.
Since the frequency of a mode is $\omega = c\,k_x$, the modes with small $k_x$ are slow, and the filter could change the
celerity of a band by favouring them. We test its strongest case, in which the fluid exchanges a given flux with the
wall, whatever the temperature of the wall; the filter coupled to the fluid is weaker (appendix~\ref{app:coating}). Each
mode of both frames is multiplied by the ratio of temperature to flux at the surface of the paint \citep{winfree2017},
\begin{equation}
  H = \frac{\tanh(\gamma d_c)}{k_s\gamma}, \qquad
  \gamma^2 = |\mathbf{k}|^2 + \frac{\omega_d + \mathrm{i}\,\omega}{\alpha_s},
  \label{eq:wallfilter}
\end{equation}
with $d_c$, $k_s$ and $\alpha_s$ the thickness, conductivity and diffusivity of the paint, and $\omega_d$ the rate at which
the mode loses coherence. One frame with a contrast $c_5/c_1 = 2$ is tracked with the phase tracker, with the loss of
coherence and the noise of the TSP dataset (\S\ref{sec:record}), in three realisations. At $\lambda = 16$~px the P\'eclet
number $c\lambda/(2\pi\alpha_s)$ goes from $10^2$ to $10^4$ and $d_c/\lambda$ from $0.013$ to $0.052$. In pixels, and in
$d_c/\lambda$, the three bands of the TSP dataset are b2, b3 and b4 of the test.

Without noise the filter changes the celerity by at most one per cent on b2 and b3, two on b1, four on b4 and eight on b5.
With noise the changes reach five per cent on b2 to b4 and about ten on b1 and b5, and are comparable with the spread of
the realisations except on b1. The test has a limit: each mode keeps its speed, so the filter only changes the weight of
the modes. In a pattern of finite life each wavenumber contains a range of frequencies, and the filter favours the slow
ones: for a life of $70$~ms the same filter removes up to seventy per cent of the celerity (appendix~\ref{app:coating}).

\subsection{Noise correlated by a temporal filter}\label{sec:bench-noise}
This test checks the prediction of \S\,\ref{sec:eq-noise}: the bias has the sign opposite to the correlation $\rho(\tau)$
of the noise, and vanishes once the interval exceeds the temporal filter. The intensity tracker is run on three pairs of
frames of the channel, with white noise passed through the temporal filter of the TSP chain (\S\ref{sec:tsp}), for which
$\rho = 0.69$, $0.15$, $-0.08$, $-0.04$ at one to four frames and negligible beyond. The noise makes $0.66$ of the streamwise
gradient energy of b3, the share measured on the finest band of the TSP dataset.

First, the second frame is the first shifted rigidly by $1.07$~px, so that only $\rho(\tau)$ changes. With the noise of the
two frames independent, the bias of the broadband median is below $5\times10^{-4}$~px. At one and two frames the
displacement is pulled towards zero, by $1.06$ and $0.96$~px; at three and four it is pushed forward, by $0.62$ and
$0.40$~px; from five frames on the bias is at most $0.03$~px. The minimum of the mean cost \eqref{eq:ssd-mean}, computed
without the tracker, moves in the same sense, so the bias comes from the cost and not from the solver.

Then the second frame is a later frame of the simulation. The finest band returns about a hundredth of its clean $\CU$ at
one frame and nearly twice it at three; from five frames on every band is within four per cent of it. The clean $\CU$ itself
changes by less than one per cent up to eight frames ($\delta t^+ = 2.6$); at the scales of the ramp bands,
$\lambda_s^+ = 214$ to $866$, it changes by at most three per cent up to $\delta t^+ \approx 15$ and seven up to $30$.

\subsection{A reference PIV code based on correlation peak}\label{sec:pivlab}
PIVlab~3.14 \citep{thielicke2014,thielicke2021}, an established correlation code, is run on the frames of three of the
tests, broadband and band-passed, to see which findings depend on the estimator. It is a baseline and not a measure of
accuracy. Three arms are used. The first has the settings of the example script: contrast equalisation, three passes of
$64$, $32$ and $16$~px with half overlap, linear window deformation, a three-point Gaussian fit of the peak and the
standard validation. The second repeats the last pass until the update falls below $10^{-3}$~px, an option of the code. The
third, also converged, sets the final window to twice the wavelength of each band, but not below $16$~px.
Table~\ref{tab:pivlab} gives the results.

\begin{table}
\centering
\small
\setlength{\tabcolsep}{4pt}
\begin{tabular}{lccc}
\toprule
 & uniform & $c_5/c_1$ & noise bias [px] \\
 & shift & (imposed $2.0$) & 1/3/5 frames \\
\midrule
intensity tracker          & $<10^{-4}$ & 1.78 & $-1.06$/$+0.62$/$0.00$ \\
phase tracker              & $<10^{-4}$ & 1.80 & ---                    \\
PIVlab, example settings   & $-15\%$    & 0.48 & $-1.05$/$+0.70$/$-0.03$ \\
PIVlab, converged          & $-4\%$     & 0.68 & $-1.05$/$+0.69$/$+0.01$ \\
PIVlab, converged, window $2\lambda$ & $-4\%$ & 1.45 & $-1.05$/$+0.69$/$+0.01$ \\
\bottomrule
\end{tabular}
\caption{The benches through a reference correlation code. Uniform shift:
error on $1.07$~px. Contrast: as table~\ref{tab:mc_kw}. Noise: broadband
bias of the median, correlated MaxPol noise (\S\ref{sec:bench-noise}).
A dash: not computed.
}
\label{tab:pivlab}
\end{table}

\emph{Uniform shift.} The example settings return fifteen per cent less than the imposed displacement. On a smooth field the
peak is broad, each pass of the deformation removes only part of the residual displacement, and three passes are not
enough. Repeated to convergence, the code is within four per cent; both trackers are within $0.005\%$
(\S\ref{sec:uniform}).
\begin{table}
	\centering
	\small
	\begin{tabular}{p{5.0cm}p{3.6cm}p{5.2cm}}
		\toprule
		choice & test & result \\
		\midrule
		sum of squared differences, continuous sub-pixel solution, deformation from the window
		& uniform shift (\S\ref{sec:uniform}, \S\ref{sec:pivlab})
		& error $< 10^{-4}$; correlation code $-15$ to $-4$~per~cent \\
		band fixed in wavelength, the same points in every band
		& imposed contrast $2.0$ (\S\ref{sec:mcdecay}, \S\ref{sec:pivlab})
		& $1.78$ and $1.80$; correlation code $0.68$ to $1.45$ \\
		window radius per band
		& sweep (\S\ref{sec:radius})
		& no cost on a uniform celerity, within $0.04$~per~cent \\
		window at the feature, affine warp
		& celerity varying along the wall (\S\ref{sec:warp})
		& placed within a few tenths of a pixel \\
		interval of the pair longer than the temporal filter
		& correlated noise (\S\ref{sec:bench-noise})
		& bias $-1.06$, $+0.62$, $0.00$~px at $1$, $3$, $5$ frames; the same for the correlation code \\
		two formulations, intensity and phase
		& stress and pressure of the channel (\S\ref{sec:dns})
		& medians of $C$ within one per cent \\
		\bottomrule
	\end{tabular}
	\caption{The choices of the trackers, the test of each and its result.}
	\label{tab:choices}
\end{table}

\emph{Contrast.} A window shorter than the wavelength sees too little of a coarse band, and its celerity falls towards
zero. With the window at $16$~px the coarsest band loses two thirds of its celerity and an imposed contrast of two is
returned as $0.68$; with a window of twice the wavelength the loss is more than a quarter and the contrast $1.45$, against
$1.78$ and $1.80$ of the trackers (table~\ref{tab:mc_kw}).

\emph{Correlated noise.} With the noise of \S\ref{sec:bench-noise} PIVlab is pulled back by $1.05$~px at one frame, pushed
forward by $0.69$~px at three and unbiased, within $0.05$~px, from five on, as the intensity tracker is. The bias comes from
the filtered frames and not from the tracker: the interval of the pair must exceed the filter also for a correlation code.

\subsection{Real evolution: the stress and the pressure of the channel}\label{sec:dns}
\begin{figure}
	\centering
	\includegraphics[width=.85\textwidth]{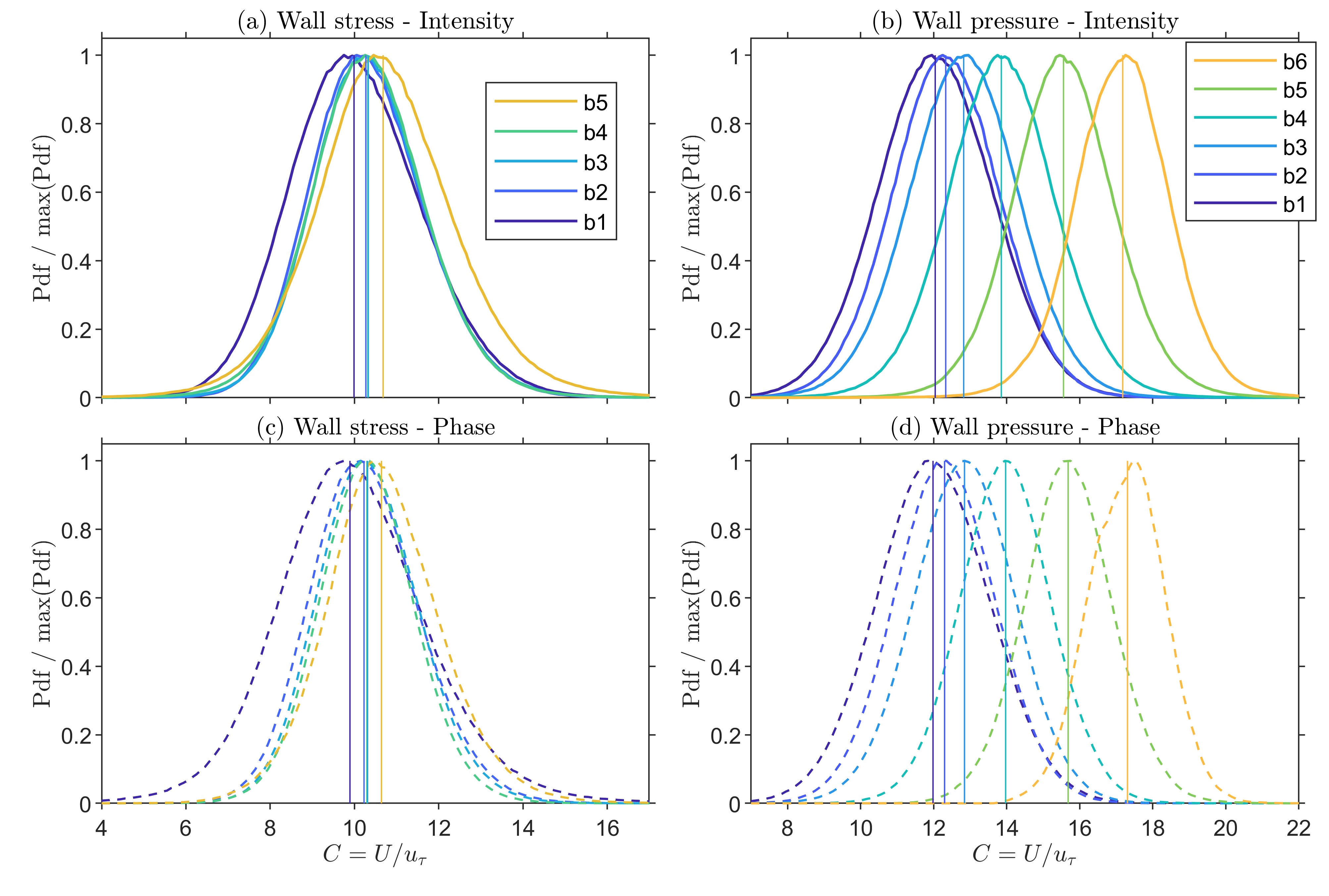}
	\caption{Probability density of the coefficient $C = U/\utau$ per band on the turbulent channel, each curve divided by
	its maximum: (a, c) wall stress,
	b1--b5; (b, d) wall pressure, b1--b6. (a, b) Intensity tracker; (c, d) phase tracker. Vertical lines, the median of
	each band; single features on the $124$ pairs of \S\ref{sec:fields}.}
	\label{fig:cpdf}
\end{figure}

A tracker gives the distribution of the celerity over time and space, and not only its mean. Figure~\ref{fig:cpdf} shows it for $C = U/\utau$
per band, for the stress and the pressure of the channel, on the $124$ pairs of \S\ref{sec:fields}. The stress pattern
travels at nearly the same speed on every band: the median of $\CU$ is $10.0$ to $10.7$ with the intensity
tracker, and half of the features lie in a range $1.8$ to $2.1$ wide around it. The pressure pattern travels faster, and faster on coarser bands: the median of $C_p$ grows from $12.0$ on b1 to
$17.2$ on b6, and the range is $1.7$ to $2.1$ wide.

The two formulations return the same body of the distribution: the medians agree within one per cent on
every band of both fields, and the quartiles within $0.3$. The difference does not come from the features that the phase
tracker rejects: on the features that both accept, the intensity median moves by at most $0.2$ per cent. It is small but
systematic, $-1.0$ to $+0.9$ per cent, with $95\%$ intervals of $0.1$ to $0.9$ points from a bootstrap over $32$ tiles of the
plane that follow the pattern; only on b5 of the stress does the interval include zero. The medians themselves are known within
$0.6$ to $1.1$ per cent. Over $1300\,t^+$, the medians of eight intervals of $162\,t^+$ differ with a standard deviation of
at most $0.4$ per cent, $0.7$ on b6 of the pressure, so the first interval, used here, represents the whole record.

The phase tracker cuts the slow tail of the coarse pressure bands, and is broader than the
intensity tracker on b1 of the stress. It loses $3$, $3$, $4$, $8$ and $15$ per cent of the features on b1 to b5, and $28$ on
b6 of the pressure; the intensity tracker loses about $0.1$ per cent at most. But the intensity tracker keeps a few features
in a hundred thousand whose solve has diverged, with celerities tens to thousands of times the mean, which the
forward--backward test does not catch; the tests of the phase tracker (\S\ref{sec:phaserej}) leave almost none. They do not
move a median.

\begin{figure}
	\centering
	\includegraphics[width=\textwidth]{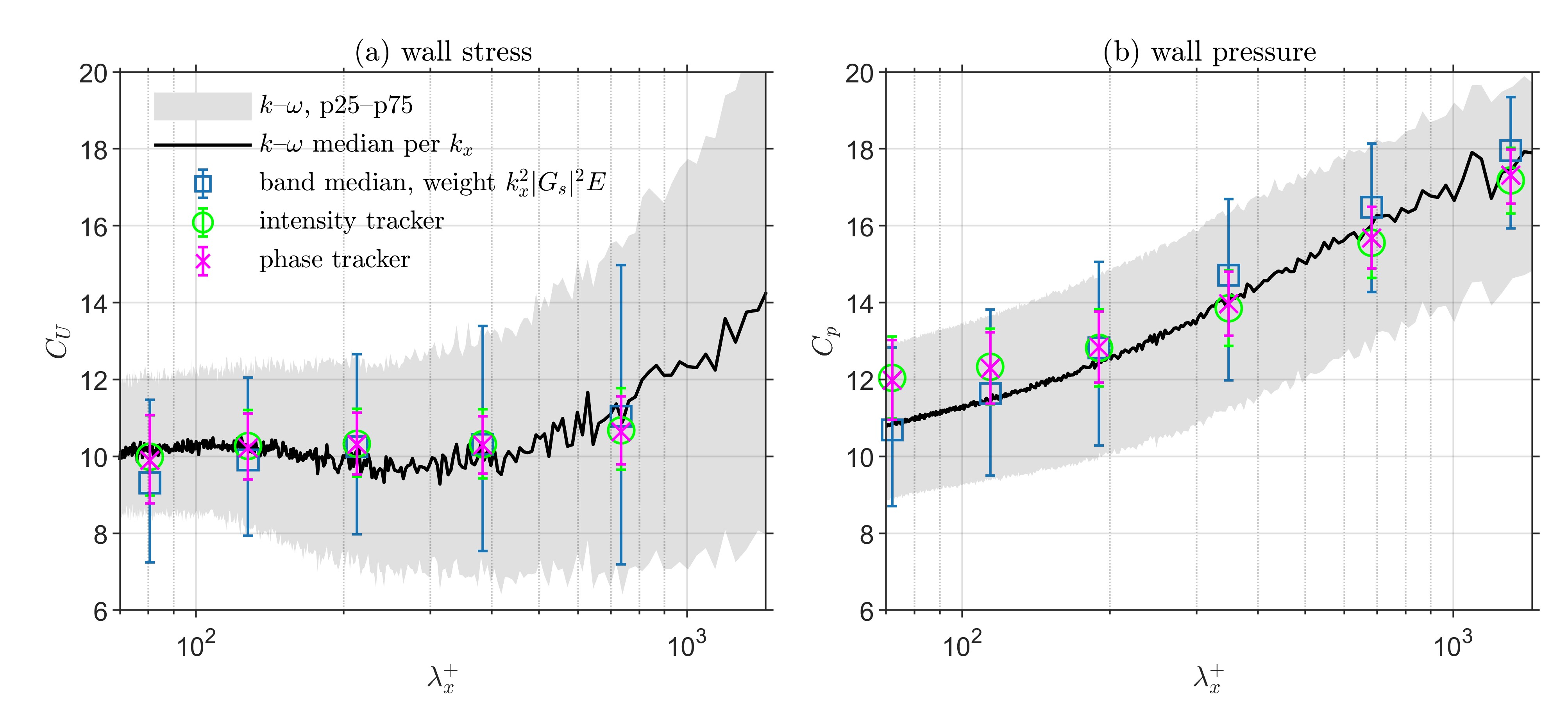}
	\caption{The coefficient $C$ against the streamwise wavelength on the turbulent channel: (a) wall stress, (b) wall
	pressure. Line and shading, median and 25--75\,\% range of the $k_x$--$\omega$ energy at each $k_x$
	\citep[][figure~2b]{miozzi2026part1}, on $500$ snapshots for the stress and $1000$ frames for the pressure
	(\S\ref{sec:fields}); squares, the same statistic over each band with the weight $k_x^2|G_s|^2E$ of
	\eqref{eq:weight}; open circles and crosses, the medians of the intensity and of the phase tracker; bars,
	25--75\,\% ranges. The bands are drawn at the streamwise
	wavelength $2\pi/\langle|k_x|\rangle$ under the same weight.}
	\label{fig:clambda}
\end{figure}

Figure~\ref{fig:clambda} places the medians against the streamwise wavelength. The line
is the median of the $k_x$--$\omega$ energy at each $k_x$, the statistic of \citet[][figure~2b]{miozzi2026part1}; the squares
are the same median over each band, with the weight $k_x^2|G_s|^2E$ of the gradient energy (\S\ref{sec:eq-weak}). On the
stress the trackers follow the squares within seven per cent on b1 and four on the other bands. On the pressure they read the
finest band thirteen per cent higher and the second six, agree on the third, and read the three coarsest four to six per cent
lower. A tracker gives one vote to each feature, while the spectrum weights each Fourier mode by its energy. The
range that holds half of the spectral weight is two to four times as wide as the one that holds half of the features.

Table~\ref{tab:choices} summarises the choices of the trackers, the test of each and its result.

\section{The thermal footprint of an attached boundary layer}\label{sec:tsp}

The trackers are applied to the upstream dataset of \S\ref{sec:record}, at the reference position, $x = -97.5$~mm, where the
boundary layer is attached. No independent measurement of the celerity exists here: the test is one of consistency, between
the formulations and between the intervals.

\begin{figure}
	\centering
	\includegraphics[width=\textwidth]{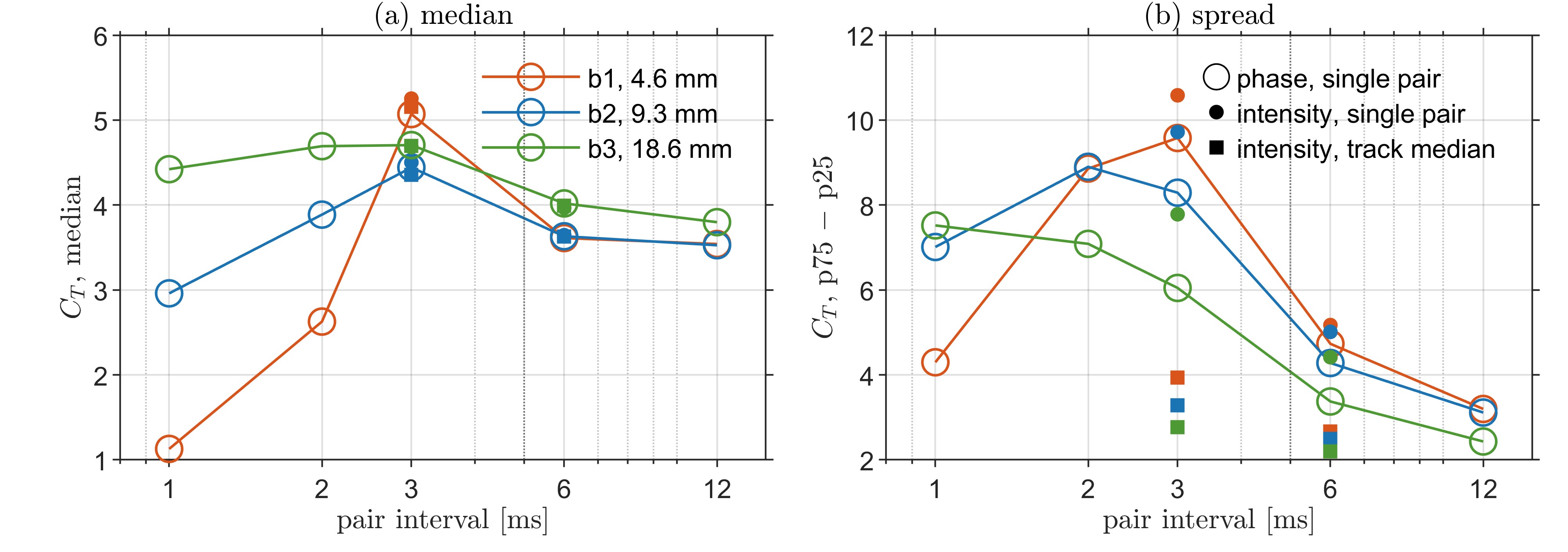}
	\caption{(a) The median and (b) the interquartile range of $\CT$ at $x = -97.5$~mm against the interval of the
		pair. Open circles and lines, phase tracker, single pair, on the first $1.8$~s; filled circles, intensity tracker,
		single pair; squares, intensity tracker, track median. Dotted, five frames, from which the noise of the two frames
		is independent.}
	\label{fig:ctsp_interval}
\end{figure}

The images pass through the noise removal chain of \citet{miozzi2019} and \citet{miozzi2020}, with a
spatial and a temporal filter. The temporal filter, a seven-tap, zero-phase low-pass of the MaxPol family
\citep{hosseini2017finite} with its half-power point at $205$~Hz, correlates the noise up to four frames apart
(\S\,\ref{sec:eq-noise}). Figure~\ref{fig:ctsp_interval} shows the consequence, with the phase tracker: on the finest band
the median of $\CT$ is $1.1$ at $1$~ms and $5.1$ at $3$~ms, $40$ per cent above its value at $6$~ms; the coarser bands are
about a fifth too high at $3$~ms. The filter no longer correlates the noise of frames five or more apart, and the
bias should vanish there. It does: from $6$ to $12$~ms $\CT$ falls by only two, three and six per cent on the three bands.
Celerities are therefore taken at $6$~ms, where the two formulations agree within two
per cent.

The filter is needed, because the noise of a single frame is comparable with the displacement. Without it, on $1$~s of
the same recording, the error of a pair hardly depends on the interval, and the spread of $\CT$ at $6$~ms is $1.3$ to $1.6$
times the filtered one. The peak at $3$~ms disappears, so it belongs to the filter; the median starts high and settles only
from about $6$~ms, which \S\ref{sec:eq-noise} does not predict. An edge-preserving spatial denoising of each frame raises the
in-band signal-to-noise ratio by up to a half and leaves both results unchanged.

\begin{figure}
	\centering
	\includegraphics[width=\textwidth]{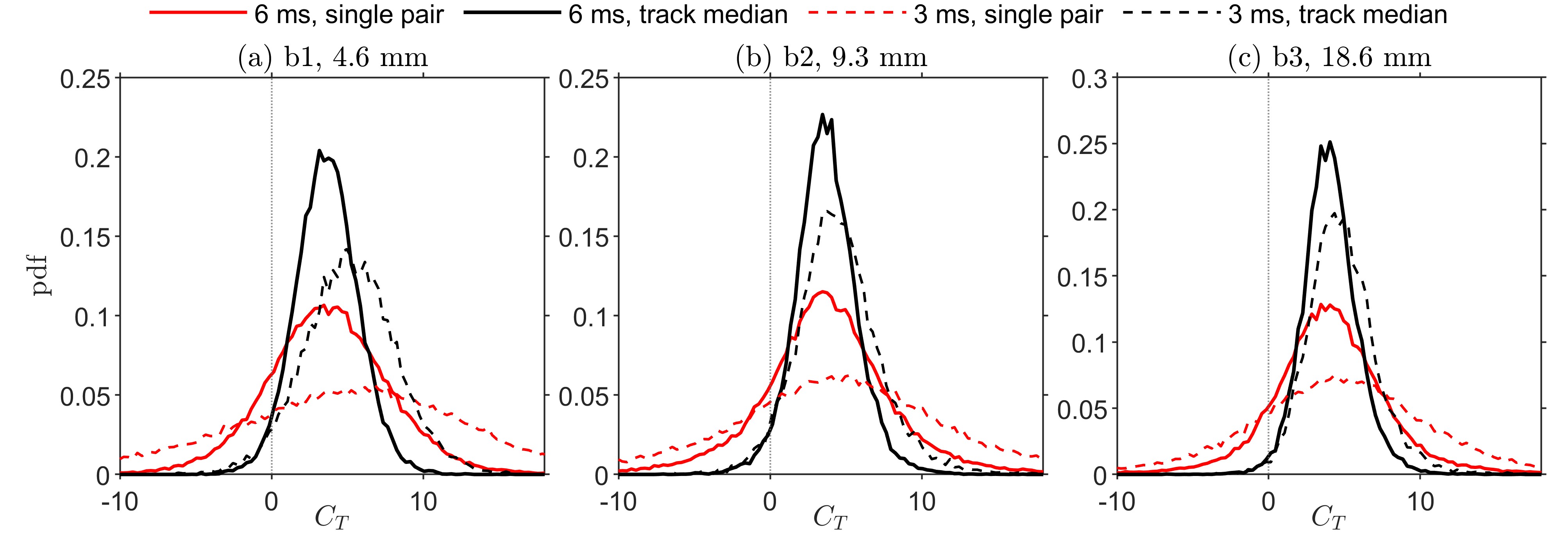}
	\caption{Probability density of $\CT$ at the reference position, $x = -97.5$~mm, on the three bands of the TSP
		dataset: single pair (gray) and track median (black), at $6$~ms (solid) and $3$~ms (dashed); intensity tracker.}
	\label{fig:ctsp_pdf}
\end{figure}

A single pair at $6$~ms scatters as much as the displacement (figure~\ref{fig:ctsp_pdf}): on the attached
wall a sixth of the samples of the finer bands point upstream, as a Gaussian of the same spread predicts. The celerity is
therefore the median of five consecutive samples of the track, $\pm15$~ms \citep[][\S II]{miozzi2026part2}. The median of
$\CT$ does not change, within one per cent; the spread halves and the upstream samples fall below four per cent. At $3$~ms the
same filter narrows the distribution but keeps the bias. Over $3.6$~s, a bootstrap over blocks of $200$~ms gives
$\CT = 3.63$, $3.63$ and $3.99$ for the three bands, with $95\%$ intervals of $\pm0.06$ to $\pm0.08$.

\begin{figure}
	\centering
	\includegraphics[width=.8\textwidth]{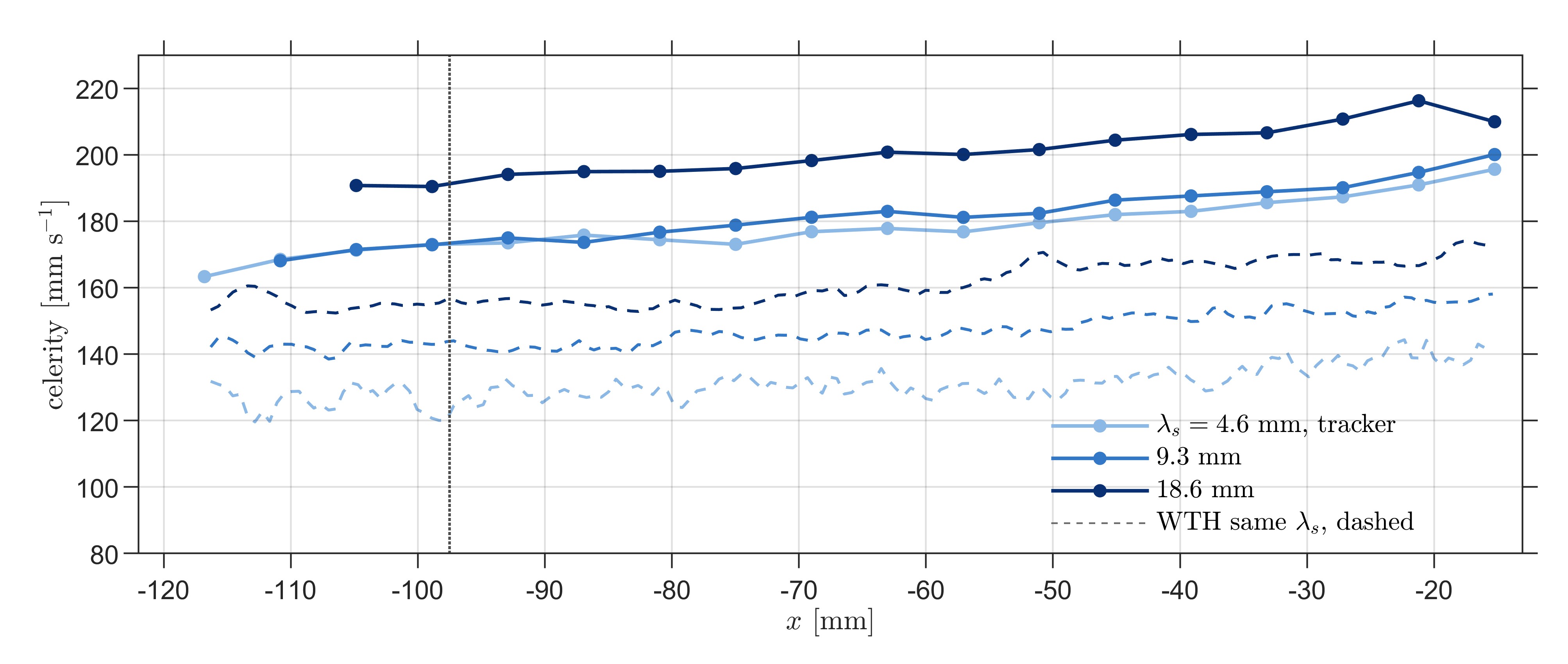}
	\caption{The celerity of the three bands on the dataset upstream of the ramp. Tracker (solid, symbols): track median at $6$~ms, median over the span in $20$ intervals of $x$. WTH (dashed, same colours): at a lag of five frames, where the noise of $A_{xx}$ vanishes, median over $|z| < 100$~mm, not drawn within $3$~mm of the upstream end of the field of view. Both are drawn up to $x = -15$~mm. The WTH curves depend on the lag (\S\ref{sec:wt-lag}). Dotted, the reference position $x = -97.5$~mm.}
	\label{fig:ramp-wth}
\end{figure}

The celerity of every band grows slowly toward the shoulder (figure~\ref{fig:ramp-wth}; the dashed
curves are those of \S\ref{sec:wt-compare}), by seven to nine per cent from the reference position to $x = -33$~mm. Part of it
is the growth of $\utau$; the rest is consistent with the thermal layer, which starts after an unheated length of $788$~mm
(\S\ref{sec:record}) and is still developing.

\section{The trackers and the weak form of Taylor's hypothesis}\label{sec:equation}

Feature tracking follows a pattern from one image to the next \citep{lk1981,baker2004}; the WTH asks, at each point, for the
velocity that best carries the whole sequence \citep{delalamo2009,miozzi2020,miozzi2024,costantini2026ramp}. They solve the
same equation, and differ in where they average and in how the noise enters.

\subsection{The weak form of the Taylor Hypothesis (WTH)}\label{sec:eq-weak}

The WTH asks for the velocity $\bm{U}$ that best carries the fluctuation
$T'$ of the wall temperature, at one point, over the whole sequence:
\begin{equation}
	\left\langle \left(\partial_t T' + \bm{U}\cdot\nabla T'\right)^2 \right\rangle = \min
	\quad\Longrightarrow\quad
	\mathbf{A}\,\bm{U} = -\bm{b} ,
	\label{eq:wt}
\end{equation}
with $A_{ij} = \langle \partial_i T'\,\partial_j T'\rangle$ and $b_i = \langle \partial_i T'\,\partial_t T'\rangle$, $i,j$ the
streamwise and spanwise directions and the average taken in time (equation~(4) of \citealp{miozzi2024}). The pattern need not
be frozen: the WTH returns the one velocity that explains most of its change. For a field homogeneous in space, stationary in
time and symmetric in $k_z$, with space-time spectrum $E(k_x,k_z,\omega)$, the streamwise component is
\begin{equation}
	U = \frac{\sum c\,k_x^2 E(k_x,k_z,\omega)}{\sum k_x^2 E(k_x,k_z,\omega)},\qquad c = \omega/k_x ,
	\label{eq:weight}
\end{equation}
a mean of the phase speeds weighted by the gradient energy. Across an octave $k_x^2$ changes by a factor of four, so the
weight favours the finest scales of a band, and we apply \eqref{eq:wt} band by band. Noise symmetric in $\omega$ adds to
$A_{xx}$ and nothing to $b_x$, and lowers $U$; a discrete derivative enters $\bm{b}$ once and $\mathbf{A}$ twice.

The tracker solves equation \eqref{eq:wt} too: its translational form (equation \ref{eq:lk}) is the same equation with the average taken over a
window at one instant instead of over time at one point, and a finite displacement instead of a derivative. The affine stage
adds the four displacement gradients, which one velocity per point does not have. The tracker meets the noise only through its
correlation between the frames (\S\,\ref{sec:eq-noise}); the WTH must use a lag at which it vanishes.

\subsection{Noise and the choice of the lag}\label{sec:wt-lag}
Equation \eqref{eq:wt} needs the space and time derivatives of the image. A difference between pixels depends on their
distance, and the celerity with it; we therefore take the space derivatives of each band in Fourier space, where there is no
distance to choose. The two derivatives are then multiplied at the same point and at the same instant. In a noisy image each
contains noise, and the noise multiplied by itself does not average to zero: it makes $A_{xx}$ too large
and the celerity too low. The remedy is to multiply derivatives taken at two different instants, $\tau$ apart, far enough that
their noise is independent:
\begin{equation}
\bm{U} = -\,\mathbf{A}(\tau)^{-1}\,\bm{b}(\tau), \qquad
A_{ij}(\tau) = \langle \partial_i T'(t)\,\partial_j T'(t+\tau)\rangle, \qquad
b_i(\tau) = \langle \partial_i T'(t)\,\partial_t T'(t+\tau)\rangle ,
\label{eq:wtlag}
\end{equation}
averaged over $\pm\tau$. The temporal filter makes the noise of nearby frames similar, up to four frames
(\S\ref{sec:eq-noise}), so we use $\tau = 5$ frames. The lag has a cost. At a fixed point the product of two instants falls also because the pattern passes, as
$\cos(kc\tau)$ for a mode of wavenumber $k$ and speed $c$, and vanishes when the mode has moved by a quarter of its wavelength.
In five frames the pattern moves by $0.19$, $0.09$ and $0.05$ of the wavelength on b1 to b3, so the estimate is least reliable
on b1. On
the turbulent channel there is no noise, and no lag is needed.

\subsection{Comparison}\label{sec:wt-compare}
The WTH is the minimum of the residual of Taylor's hypothesis. Divided by the variance of $\partial_t T'$, the minimum is
\begin{equation}
\epsilon = 1 - \frac{\bm{b}^{\mathsf T}\mathbf{A}^{-1}\bm{b}}{\langle(\partial_t T')^2\rangle} ,
\label{eq:eps}
\end{equation}
the part of the change in time that does not obey the hypothesis, whatever the constant velocity: $\epsilon = 0$ for
a pattern that moves at one speed without changing, and $1$ for one that no speed describes. 

On the channel the WTH and the two trackers agree within three per cent on b2 to b4, and within seven on b1 and b5
(table~\ref{tab:dns}). Two methods with different averages, one over time at a point and the other over a window at an instant,
return close values of the celerity, although the WTH varies more from band to band. One celerity accounts for $77$ to $88$ per cent of the change ($\epsilon = 0.12$ to $0.23$). Part of
the remaining difference is the weight, which in the WTH favours the steep gradients: applied to the tracker, it recovers
between a tenth and two fifths of it, on the four bands where the two differ.

\begin{table}
\centering
\begin{tabular}{lccccc}
\toprule
band & $\lameff^+$ & WTH & intensity tracker & phase tracker & $\epsilon$ \\
\midrule
b1 & 56  & 9.32  & 9.99  & 9.89  & 0.16 \\
b2 & 97  & 9.99  & 10.27 & 10.23 & 0.12 \\
b3 & 173 & 10.35 & 10.33 & 10.31 & 0.12 \\
b4 & 341 & 10.58 & 10.32 & 10.30 & 0.16 \\
b5 & 705 & 11.34 & 10.68 & 10.64 & 0.23 \\
\bottomrule
\end{tabular}
\caption{The coefficient $\CU = U/\utau$ on the turbulent channel (stress
pattern at $y^+ = 0.22$, all $500$ stored frames): WTH
\eqref{eq:wt} on the plane-averaged moments, the intensity and the phase tracker on the same features and
windows, and
$\epsilon$ of
\eqref{eq:eps}. $\lameff^+$ is the effective radial wavelength of the band \citep[][table~1]{miozzi2026part1}; the axis of
figure~\ref{fig:clambda} is the streamwise one.
}
\label{tab:dns}
\end{table}

On the thermal footprint (\S\ref{sec:record}, figure~\ref{fig:ramp-wth}) the noise requires the lag. At the reference position
the WTH gives $120$, $144$ and $156$~mm\,s$^{-1}$ at five frames, the tracker $174$, $174$ and $191$. The WTH is lower on every
band, most on b1, where the lag weighs most; at seven frames it gives $87$, $129$ and $160$. The comparison is therefore
indicative. Both grow along the wall.

The two tools are complementary. The WTH gives the mean celerity at every point without features or windows, and $\epsilon$. The
tracker gives the celerity of each feature at each instant, and follows the pattern between the frames of a pair, so its
interval does not meet the cost of the lag; we use it for the values of this paper. A celerity should be quoted with the method,
the band and the interval or lag, and $C$ calibrated with the same ones.

\section{Designing a measurement}\label{sec:design}
The tests above fix how a measurement, or a simulation, must be laid out to return the celerity of a given range of
scales. Table~\ref{tab:design} collects the rules. The design needs a reference velocity, $\utau$ on an attached wall,
and the range of celerity of the flow, including its sign: a negative celerity constrains the interval as a positive
one does.
\section{Conclusions}\label{sec:conclusions}
The celerity of a wall footprint can be measured as a field: a value per feature, per scale and per instant, on the same wall
points for every scale. A convection velocity is usually one number, or one value per wavenumber. Here it is local in space and time, and can be
followed and mapped. This paper proposes a measurement methodology and reports its accuracy, without interpreting it physically. The link with the motions above the wall and its use in a separated flow are in the companion papers \citep{miozzi2026part1,miozzi2026part2}.
\begin{table}
	\centering
	\small
	\begin{tabular}{p{3.0cm}p{5.4cm}p{5.0cm}}
		\toprule
		quantity & rule & origin \\
		\midrule
		pixel $\Delta$ & $\Delta \le \lambda_{\min}/4$ without noise, $\Delta \le \lambda_{\min}/8$ with noise
		& channel: the $4$~px band agrees with the $k$--$\omega$ spectrum; TSP: $8$~px finest band \\
		window radius $R$ & $\approx 1.5\,\lambda_0$ at $4$~px, $\approx \lambda_0$ at $8$~px, $\approx 0.65\,\lambda_0$ above
		& \S\ref{sec:radius} \\
		field $L$ & $L \gtrsim 6\,\lambda_{\max} + 2m$, edge margin $m \approx 0.4\,\lambda_{\max}$
		& coarsest band of the channel; edge gate \\
		interval, upper bound & $\Delta t \le \lambda_{\min}/(4\,|U|_{\max})$ and $\Delta t^+ \lesssim 5$
		& displacement below a quarter wavelength; decorrelation of the fine bands \\
		interval, lower bound & $\Delta t \ge (n_c+1)/f_s$, with $n_c$ the frames over which a temporal filter correlates the
		noise; $n_c = 4$ for MaxPol & \S\ref{sec:bench-noise}, figure~\ref{fig:ctsp_interval} \\
		acquisition rate & $f_s \ge 4\,(n_c+1)\,|U|_{\max}/\lambda_{\min}$ & the two bounds together \\
		single-pair scatter & median along the track over $W$ frames; sign by $|\bar U| > 2s/\sqrt{W}$
		& figure~\ref{fig:ctsp_pdf} \\
		$k$--$\omega$ spectrum & contiguous record, $T^+~\ge~ \lambda_{x,\max}^+/(0.07\,C)$
		& \S\ref{sec:fields}: $T^+ \approx 1300$ for the pressure \\
		uncertainty & bootstrap over blocks of at least $160\,t^+$ & blocks of $80\,t^+$ remain correlated \\
		\bottomrule
	\end{tabular}
	\caption{Rules for the design of a celerity measurement or simulation. $\lambda_{\min}$, $\lambda_{\max}$: the finest
		and the coarsest band; $\lambda_0$: nominal wavelength of a band in pixels; $|U|_{\max}$: the largest magnitude of the
		celerity on the finest band; $f_s$: the frame rate; $T^+$: the duration of the record. The lower bound on the interval and the quarter wavelength follow
		from the analysis of \S\ref{sec:class}, and the acquisition rate from the two together; the other rules are empirical.}
	\label{tab:design}
\end{table}
If the celerity is spatially uniform, the trackers return it within $0.005\%$. In general, the celerity changes from one scale to another and from one point to another. The trackers return its average over the band and over the window (\S\ref{sec:mcdecay}, \S\ref{sec:warp}).

On real data the pattern also changes as it moves, so no exact celerity exists to compare with, and we compare different estimates: the two trackers with each other, and, on the channel, with the $k_x$--$\omega$ spectrum.
The intensity and the phase tracker agree within one per cent on the channel and two on the thermal dataset, so the result does not depend on the quantity
held constant. A correlation-peak code needs a window of twice the wavelength and still loses part of the coarse bands; the trackers use a smaller window on each band.

Additive noise biases the celerity through what the two frames share, and not through its level, which sets the scatter. A temporal filter makes them share it: with $M$ taps it
correlates the noise of frames up to $M-1$ apart, whatever the frame rate, so the interval of the pair, counted in frames,
must exceed this length. With the seven taps used on the thermal dataset the correlation is negligible from five frames on, and
the pairs are six frames apart. 

A uniform loss of intensity biases the trackers by at most one per cent; a loss linked to
the local stress biases mainly the coarsest band, by about a fifth when the amplitude falls by a quarter. The coating changes the celerity by less than five per cent at $145$~mm\,s$^{-1}$ and by up to thirteen at $29$~mm\,s$^{-1}$ in the coupled model
(appendix~\ref{app:coating}).

Applied to the footprints of wall turbulence, the trackers already give general facts on
how they travel, which the companion papers interpret. On the channel the stress pattern travels at nearly the same speed on every band, $C_U = 10.0$ to $10.7$, and
the pressure pattern faster on coarser bands, $C_p = 12.0$ to $17.2$. Both follow the $k_x$--$\omega$ spectrum within seven per
cent, except on the finest band of the pressure, thirteen per cent higher. On the attached boundary layer $\CT = 3.6$ to $4.0$.

The trackers and the WTH solve the same equation, which says that the pattern moves without changing, but they average it
differently: a tracker over a window at one instant, the WTH over time at one point, with more weight on the steep gradients.
If the pattern moved rigidly at one speed, the two would return that speed. A real pattern changes as it moves, and they
differ. On the channel a single velocity leaves $12$ to $23$ per cent of the change in time unexplained, and the two estimates
agree within seven per cent. The unexplained share is thus a measure of how far a footprint is from a pattern that is only
carried. On the thermal footprint the noise requires a lag, on which the WTH depends, and the WTH is lower than the tracker.

The celerity therefore depends on how it is estimated and on the scale, and so does the coefficient $C$ that turns it into a
friction velocity. A tracker keeps the celerity of every feature on every band, so it can be averaged with the weight of
another estimator, and the choice is explicit; the WTH fixes the weight in advance. On the channel the weight alone accounts
for a tenth to two fifths of the distance between the two (\S\ref{sec:wt-compare}), so $C$ must still be calibrated with the
estimator, the band and the interval or lag with which it is used.

Every point of a broadband track carries the celerity of all the scales: the footprint becomes a set of Lagrangian probes at the wall. The loss of coherence along a path and the coupling between scales on one element of the pattern can then be measured, in DNS simulations and, with paint, in TSP and PSP experiments.

\appendix
\section{The thermal response of a coating}\label{app:coating}
The wall is covered by a luminescent coating, and the camera records the temperature of the coating. The coating stores and
conducts heat, so it filters the footprint of the flow in space and time, and could displace it or change its celerity.

Let $\theta_f$ be the footprint that a wall without thermal inertia would show, and $\theta_s$ the temperature at
the surface of the coating. For a mode of wavenumber $\mathbf{k}$ and frequency $\omega$,
\begin{equation}
  \frac{\theta_s}{\theta_f} = \frac{Y_f + h}{Y_f + h + Y_w}, \qquad
  Y_f = k_f\gamma_f, \qquad
  Y_w = k_s\gamma_s \tanh(\gamma_s d_c) \;\;\text{or}\;\; k_s\gamma_s \coth(\gamma_s d_c),
  \label{eq:conjugate}
\end{equation}
where $Y_f$ and $Y_w$ are the admittances of the fluid and of the coating, $d_c$ its thickness, $h$ the mean heat-transfer
coefficient, $\gamma_m^2 = |\mathbf{k}|^2 + \mathrm{i}\,\omega/\alpha_m$, and $k_m$, $\alpha_m$ the conductivity and diffusivity
of the fluid ($m = f$) and of the coating ($m = s$). The $\tanh$ holds for a substrate with a constant flux (adiabatic for the fluctuations), the $\coth$ for an
isothermal one; a real substrate lies between the two.

If the frequency is high enough that the heat does not reach the substrate and
$|\mathbf{k}|^2 \ll \omega/\alpha_m$, and without $h$, the ratio is real, $e_f/(e_f + e_s)$, with $e_m = k_m/\sqrt{\alpha_m}$
the effusivity: the coating reduces the amplitude and does not delay it. The term $h$ brings a phase where it is comparable
with $Y_f$; near the zero of the stress it falls and the phase with it.

If the fluid imposed a given flux, whatever the temperature of the wall, the response would be $1/Y_w$: the filter $H$ of
\eqref{eq:wallfilter} without loss of coherence on an isothermal substrate, and $\coth(\gamma_s d_c)/(k_s\gamma_s)$ on one with a constant flux. On the pairs of \S\ref{sec:mccoating} the two give, to first order, the same celerity of each band,
within $0.2$ per cent for P\'eclet numbers of $10^3$ and above and within two at $10^2$. A thick layer would lag the flux by
$45^\circ$, a displacement of $\lambda/8$ upstream \citep{mosyak2001}. This limit requires $Y_f \to 0$; in water the effusivity
of the fluid is about two and a half times that of the paint, so it does not apply \citep{tiselj2001}.

In the dataset of the Backward Facing Ramp, we take water with $k_f = 0.60$~W\,m$^{-1}$\,K$^{-1}$ and $\alpha_f = 1.43\times10^{-7}$~m$^2$\,s$^{-1}$, and a paint with $\alpha_s = 10^{-7}$~m$^2$\,s$^{-1}$ and $e_s/e_f = 0.4$, both
assumed, hence $k_s = 0.20$~W\,m$^{-1}$\,K$^{-1}$. The thickness is $100$ or $240\,\mu$m, $h$ from $1500$ to $6000$~W\,m$^{-2}$\,K$^{-1}$, around the $3000$ given by the Reynolds analogy, and both substrates are used. At the frequencies of the bands the phase lies between $-8^\circ$ and $+5^\circ$: less than $0.025\,\lambda$, under $0.5$~mm on the coarsest band.

A pattern of finite life, from $70$ to $1000$~ms, has a spread of frequencies at each wavenumber, and its celerity through the filter is predicted from the spectrum of the dataset. It changes by less than five per cent at $145$~mm\,s$^{-1}$. At $29$~mm\,s$^{-1}$ it changes by up to thirteen, on the coarsest band, for the shortest life and the thinner coating. With the prescribed flux the loss reaches seventy per cent on tracked synthetic sequences, which the prediction follows within $0.04$.

\section*{Statements}
\paragraph{Acknowledgments.}
	MM gratefully acknowledges Dr. Christian Klein, Dr. Marco Costantini, Dr. Andreas Schr\"oder and the Dept. of Experimental Methods of the DLR Institute of Aerodynamics and Flow Technology for the meaningful discussions and continuous support developed in these years of fruitful collaboration.

	MM also gratefully acknowledges the Johns Hopkins Turbulence Database (JHTDB) for providing open access to their Direct Numerical Simulation dataset of the Turbulent Channel Flow at $Re_{\tau}=1000$ \citep{graham2016web}.

	MM also gratefully acknowledges Dr. Marco Costantini (Dept. of Experimental Methods of the DLR Institute of Aerodynamics and Flow Technology) and Dr. Cornelia Grabe ($C^2A^2S^2E$ Department, Center for Computer Applications in AeroSpace Science and Engineering, DLR Institute of Aerodynamics and Flow Technology) for providing the TSP dataset of the Backward Facing Ramp examined in this study.

	During the preparation of this work the author used Claude (Opus 5.5) to assist in optimizing and testing the tracking codes, to support in writing scripts for the benchmarks and the analysis, and in drafting, shortening and editing the text. 
	The original concepts, methodology, and intellectual contributions remain exclusively those of the author.
	The author reviewed and edited all the output, and takes full responsibility for the content of the publication.
\paragraph{Supplementary information}
\emph{Movie 1} shows the trajectories tracked on three bands (b1, b3, b5) of the wall stress and of the wall pressure of the turbulent channel at $\Ret = 1000$. The background is the field filtered on the band; each trajectory is coloured by the streamwise celerity of that band, $C = U/u_\tau$, and shows its last $15$ instants. The displayed area is a $256 \times 128$~px subregion of the $2048 \times 768$~px domain. About $43\,000$ trajectories are tracked at each instant on the full domain; for readability the movie shows one in four of those that cross the displayed area.
\paragraph{Funding}
This research received no specific grant from any funding agency in the public, commercial, or not-for-profit sectors. It was a curiosity-driven study conducted using institutional internal resources from CNR-INM. The TSP dataset of the Backward Facing Ramps examined in this study was obtained within the framework of the DLR ADaMant project, financially supported by the German Federal Ministry for Economic Affairs and Energy.
\paragraph{Competing interests}
The author declares no competing interests.
\paragraph{Data availability statement.}
The tracking outputs and the data behind the figures are available from the author on reasonable request.

\bibliographystyle{abbrvnat}
\bibliography{refs}

\end{document}